\documentclass{jfm}
\usepackage{graphicx}
\usepackage{epstopdf, epsfig}
\usepackage{hyperref}
\usepackage{bm}
\usepackage{color,soul}
\usepackage{mathrsfs}
\usepackage{amsmath,amssymb}
\usepackage{xcolor}

\shorttitle{Blood drop evaporation}

\shortauthor{Roy et al.}

\title{Insights into the mechanics of pure and bacteria-laden sessile whole blood droplet evaporation}

\author{Durbar Roy\aff{1}\thanks{The authors contributed equally},
Sophia M\aff{2}\footnotemark[1],
Kush K Dewaangan\aff{2},
Abdur Rasheed \aff{2},
Siddhant Jain \aff{2},
Anmol Singh\aff{3},
Dipshikha Chakravortty \aff{3},
\and Saptarshi Basu\aff{2}
  \corresp{\email{sbasu@iisc.ac.in}}}

\affiliation{
\aff{1} International Centre for Theoretical Sciences, Tata Institute of Fundamental Research, Bengaluru, 560089, India
\aff{2} Department of Mechanical Engineering, Indian Institute of Science, Bengaluru, 560012, India
\aff{3} Department of Microbiology and Cell Biology, Indian Institute of Science, Bengaluru, 560012, India\\

}

\begin{document}

\maketitle

\begin{abstract}
We study the mechanics of evaporation and precipitate formation in pure and bacteria-laden sessile whole blood droplets in the context of disease diagnostics.
Using experimental and theoretical analysis, we show evaporation process has three stages based on evaporation rate. In the first stage, edge evaporation results in a gelated contact line along the periphery through sol-gel phase transition. 
The intermediate stage consists of gelated front propagating radially inwards due to capillary flow and droplet height regression in pinned mode, forming a wet-gel phase. We unearthed that the gelation of the entire droplet occurs in the second stage, and the wet-gel formed contains trace amount of water. In the final slowest stage, wet-gel transforms into dry-gel, leading to desiccation-induced stress forming diverse crack patterns in the precipitate.
Slow evaporation in the final stage is quantitatively measured using evaporation of trace water and associated transient delamination of the precipitate.
Using axisymmetric lubrication approximation, we compute the transient droplet height profile and the erythrocytes concentration for the first two stages of evaporation. 
We show that the precipitate thickness profile computed from the theoretical analysis conforms to the optical profilometry measurements.
We show that the drop evaporation rate and final dried residue pattern do not change appreciably within the parameter variation of the bacterial concentration typically found in bacterial infection of living organisms. However, at exceedingly high bacterial concentrations, the cracks formed in the coronal region deviate from the typical radial cracks found in lower concentrations.
\end{abstract}

\begin{keywords}
Authors should not enter keywords on the manuscript, as these must be chosen by the author during the online submission process and will then be added during the typesetting process (see http://journals.cambridge.org/data/\linebreak[3]relatedlink/jfm-\linebreak[3]keywords.pdf for the full list)
\end{keywords}

\section{Introduction}
In this study, we investigate the mechanics of sessile 
droplet evaporation in pure and bacterial laden blood droplets.
Understanding the behavior of blood droplet evaporation is essential in forensic science, biotechnology, and the development of advanced medical diagnostics
\citep{chen2016blood,cameron2018biofluid,kokornaczyk2021diagnostic,pal2020concentration,pal2021temperature,pal2023drying,seyfert2022stability,demir2024artificial,brutin2015droplet,brutin2018recent,brutin2022drying}.
In particular, forensic investigators routinely uses blood stain and blood splatter pattern analysis to recreate and understand a crime scene in exclusive detail \citep{hulse2005deducing,benabdelhalim2022drying}. Blood splatter and corresponding stains are caused due to a myriad of blood droplets impinging various surfaces and interfaces \citep{roy2019dynamics,roy2022droplet,roy2023mechanics,yarin2017collision}. On equilibrium, blood drops and films becomes sessile and subsequent evaporation leads to residues which forms the basis of forensic investigation. The mechanics of drop impact \citep{yarin2017collision} and evaporation of thin films/drops
\citep{sultan2005evaporation,brutin2015droplet}
becomes important for such forensic investigations in general.
We also observe evaporation in various biological and bodily fluids like blood clots, desiccation in wounds \citep{sobac2011structural,brutin2011pattern,laux2016ultrasonic}, biofilm formations \citep{wilking2013liquid}, and drying of tears \citep{traipe2014dynamics,roy2024future} to name a few that are found in living systems.
Apart from forensics and disease diagnostics, analysis of desiccated biological samples is essential in several applications like archaeological studies \citep{wilson2013archaeological,kooyman1992verifying}, micro-array applications for proteins, nucleic acids (DNA, RNA), genotype and phenotype studies \citep{blossey2002contact,dugas2005droplet,smalyukh2006structure}. Analyzing the characteristics of sessile blood droplets during and after evaporation can provide valuable insights into the blood sample's source, age, and composition.
The physics of sessile blood drop evaporation falls at the intersection of various disciplines like fluid mechanics, solid mechanics, soft matter, statistical physics, surface chemistry, colloidal physics, and biology, to name a few \citep{brutin2015droplet}; and as a result quantitative understanding of blood droplet evaporation and its corresponding precipitate formation mechanics still remains in its infancy \citep{lanotte2017role}.

Blood is a complex naturally occurring fluid consisting of formed cellular elements suspended in blood plasma and are found in all humans and complex animal species \citep{baskurt2003blood,hoffbrand2019hoffbrand}. Various cellular elements like the biconcave (at rest) erythrocytes (red blood cells (RBCs)), formless leukocytes (white blood cells (WBCs)), and  disc shaped thrombocytes (platelets) form approximately 45-50\% of the blood volume, with RBCs having the most significant percentage among the cellular elements. On the other hand, blood plasma is an amber color liquid component of the blood, which makes approximately 50-55\% of the total blood volume. 94-95\% of the blood plasma by volume is essentially water, with the remaining 5-6\% being dissolved proteins (fibrinogen, albumin), electrolytes, glucose, clotting factors, hormones, carbon dioxide, and oxygen.
As discussed above blood is a highly concentrated colloidal suspension and its concentration is usually described by the
volume fraction/Hematocrit ($Ht$). 
For healthy individuals the physiological range of hematocrit is $Ht{\sim}40-45\%$ \citep{reinhart2016optimum}.
The very high concentration of RBCs makes optical diagnostics of evaporating blood droplets challenging in general. Based on volume ratio, blood can be thought of comprising of two major components, RBCs ($V_{RBCs}/V{\sim}40-45\%$) and plasma ($V_{plasma}/V{\sim}60-55\%$) where $V_{RBCs}$, $V_{plasma}$ represents the respective volumes in a total blood volume $V$. We can, therefore, approximate blood as a colloidal suspension \citep{israelachvili2011intermolecular} of RBCs in plasma to first-order accuracy. The morphology of dried residues of blood drops hence depends primarily on the deposition of RBCs, the major solute component of blood plasma. However, other cellular components of blood (leukocytes and thrombocytes), dissolved proteins, electrolytes, hormones, and gases have a secondary role in the dried residue pattern formation. Such secondary effects are essential in developing diagnostic solutions for particular diseases in general \citep{chen2016blood,kokornaczyk2021diagnostic}. However, in this work, we focus primarily on the deposit pattern caused by RBCs. 

Evaporation of biofluids like blood, saliva, mucus, and urine has been explored majorly in sessile droplet and thin film configurations on solid substrates due to the simplicity of the experimental setup \citep{hu2002evaporation,hu2005analysis,brutin2015droplet,kokornaczyk2021diagnostic,wilson2023evaporation}. The major observations for sessile blood droplet evaporation were the formation of a ring-like deposit at the contact line and a complex set of crack patterns in the desiccated drops and films. The different patterns formed in evaporating drops and films of biological fluids resembled patterns found in various colloidal systems hinting towards some unifying underlying laws that were common to biofluids, colloids, and pure fluids in general
\citep{deegan1997capillary,deegan2000contact,dufresne2003flow,allain1995regular,bhardwaj2009pattern,bhardwaj2010self,tarasevich2011desiccating,denkov1992mechanism,adachi1995stripe,stauber2014lifetimes,kaplan2015evaporation,chen2016blood}.
Previous studies by scientists and engineers using model fluids like mucin revealed that both deposition and crack formation depends strongly on the particle concentration and types of dissolved salts, macro-molecules, proteins, and surfactants \citep{nguyen2002patterning}.
This work explores blood drop evaporation physics in depth and also tests whether microbial bacteria present in blood simulating bacterial infection found in living organisms can affect the evaporation process and the resulting dried precipitate.  
  
 It has been shown by various research groups \citep{yakhno2005drying,shatokhina2004bio,esmonde2014characterization,kokornaczyk2021diagnostic} that pathological changes of solute and solvent composition affect the morphology of the dried residues in evaporating drops and films. Up to the recent decade, desiccation studies on biological fluids have majorly focused on solutions, but the evaporation physics of drying suspensions is relatively sparse and qualitative in nature. Some of the recent seminal work of Brutin et al. \citep{brutin2011pattern}, Chen et al. \citep{chen2016blood} and Pal et al. \citep{pal2020concentration,pal2021temperature,pal2023drying} 
being an exception. Brutin and his collaborators \citep{brutin2011pattern} analyzed blood drop evaporation mechanism, desiccation time, and dried precipitate morphology using optical diagnostics and mass measurements. 
Further, Brutin et al. \citep{brutin2011pattern} had provided five stages of blood droplet evaporation based on qualitatively observed features. The stage classification were chosen based on visual cue and hence lack key quantitative insights.
In principle, the stages can be reduced into fewer stages (2-3) based on quantitative data as was shown by Sobac et al. \citep{sobac2011structural}. Sobac and Brutin in 2011 \citep{sobac2011structural} showed that the blood droplet evaporation can be divided into two major stages overlapping with a transition stage which essentially makes it a three stage process and the current work follows similar reasoning.

In this work we follow a similar 3 stage classification based on quantitative droplet regression data. In the present work the major stages naturally emerge and can be uniquely determined from the normalized droplet volume regression curve (refer to figure 2(b)). The first stage of evaporation  (stage A) ends when the evaporation curves deviate from the initial linear evolution. The end of second stage and the transition regime encompasses the second stage (stage B) and the final stage (stage C) represents the final and very slow evaporation rate where processes like major crack formation and precipitate delamination occurs.
Although the work of Brutin et al. being highly important, as it outlines the essential physics there exists some fundamental shortcomings related to the mechanistic understanding of the underlying processes involved. One important example being the cause of the internal flow generated during the evaporation of blood droplets. In 2011, Brutin et al. reasons that the internal flow is due to Marangoni effects owing to the small value of Reynolds, Capillary and Rouse number. However as was shown later \citep{chen2016blood,lanotte2017role,iqbal2020understanding}, the internal flow generated during evaporation is essentially capillary flow which produces coffee ring kind of deposition in pure/dilute colloidal solutions/suspensions as was shown by the seminal work of Deegan \citep{deegan1997capillary,deegan2000contact}. Further the circular spots at the end of evaporation was associated to drying spots in Brutin et al. 2011 \citep{brutin2011pattern}; however as was conjectured later in 2014 by Sobac and Brutin \citep{sobac2014desiccation} the circular spots essentially indicates laminated regions (regions have relatively higher water content) and surrounding delaminated regions. 
Recently Pal et al. in a series of work \citep{pal2020concentration,pal2021temperature,pal2023drying} provided important insights related to phase transition and the role of blood proteins on the precipitate pattern in dilute blood solutions. Majority of the previous studies \citep{iqbal2020understanding,chen2016blood,lanotte2017role,du2022internal,pal2020concentration,pal2021temperature,pal2023drying} are based on diluted blood solution, however there is a need to investigate whole blood droplet evaporation from a mechanistic perspective in order to test the feasibility of various blood precipitate bio-markers as diagnostic tools \citep{chen2016blood,trantum2012biomarker}. Further, owing to a very high $Ht$, investigating whole blood droplet evaporation is challenging both experimentally \citep{bodiguel2010imaging,lanotte2017role} and theoretically. We therefore observe that there is a need for a better mechanistic and quantitative measurements/understanding of the various processes that occurs during whole blood droplet evaporation in the context of various disease diagnostic.

The current work studies blood droplet evaporation and provides insights into the physics of evaporation and precipitate formation in bacteria-laden whole blood droplets using high fidelity optical diagnostics and theoretical analysis.
We also investigate the effect of bacterial concentration on evaporation, precipitate and crack formation. The spatio-temporal evaporation and precipitate dynamics in first two stages (A, B) is modeled through coupling between evaporation flux, droplet height and RBCs  concentration from first principles using axisymmetric lubrication model.
Further the droplet precipitate is studied using micro/nano characterization techniques like profilometry, scanning electron microscope (SEM), and optical microscopy that provides additional quantitative insights into the physics of precipitate formation, RBCs distribution.
The wet-gel to dry-gel transition during the very slow stage C of evaporation is measured quantitatively using trace water evaporation and precipitate delamination dynamics.
The bacterial motion and distribution is further mapped using live fluorescence confocal microscopy. In general, the evaporation and the dried residue of a desiccating biological fluid such as blood depend on the ambient environmental factors (like temperature and humidity), substrate characteristics (like chemical composition, wettability, adhesion), droplet geometry, fluid mechanical (like density, viscosity, surface tension, pathology), physicochemical, thermofluidic, and biophysical properties. The dried blood drop dominant pattern consists of a thick rim called a corona and a center region of relatively small thickness. The coronal rim thickness is caused due to the transport of RBCs by capillary flow induced by sessile drop evaporation \citep{hertaeg2021pattern,chen2016blood,sobac2011structural,sobac2014desiccation,brutin2015droplet}. The interaction of desiccation induced stress and thickness variation from the drop center to the outer rim region in the radial direction causes crack patterns of different morphologies. Radial cracks are formed in the rim region, whereas mud-flat cracks are formed in the center and outer periphery of the dried residues. Evaporating colloidal and polymeric sessile drops analogously form a skin layer at the air-liquid interface, which eventually undergoes a buckling. Current literature on blood drop evaporation is primarily based on observational characteristics of drop shape, mass, and crack patterns \citep{hertaeg2021pattern,choi2020crack,mukhopadhyay2020interfacial,iqbal2020understanding,bahmani2017study}. However, a fundamental mechanistic understanding of the blood drop desiccation process is far from complete and requires thorough quantitative analysis. This work serves as a bridge to understand the mechanics of blood drop evaporation using ideas from colloidal desiccating drops. We decipher the various stages and the allied mechanisms of blood drop drying using high-fidelity color optical imaging, quantitative theoretical modeling, and experimental micro/nano-characterization. It is important to note that the current work establishes the mechanics
/mechanisms of the various processes and stages related to
blood droplet drying and precipitate formation physics in evaporating
pure and bacteria laden sessile blood droplet in a quantitative manner
using experimental and theoretical methods. The detailed experimental
characterization of various physical parameter ranges is outside the
scope of the present work. Further, to the best of our knowledge, this
study is the first of its kind studying the effect of bacteria on
evaporation and precipitate formation in sessile droplets in a
comprehensive manner using both experimental and theoretical methods.
It is important to note that the theoretical framework developed in
this work is based on the scaling of the governing conservation laws
and the corresponding experimental data. The motivation of the
theoretical framework was to establish the time varying droplet
thickness profile, final precipitate deposit profile, RBCs
concentration, gelation front propagation, and a physical
understanding of the crack size and flake size based on the
precipitate thickness characteristics. The exact analytical or
numerical solutions of the fully coupled unsteady 3D governing
equations of mass, momentum, energy and concentration is outside the
scope of the present study.
The authors want to highlight that the current manuscript comprehensively
investigates the \textit{coupled evaporation-precipitation mechanics} and the
\textit{various stages} of \textit{bacteria laden} sessile \textit{whole blood droplet}
evaporation from \textit{rigorous first principles} using both \textit{experimental} and \textit{theoretical analysis},
which does not exist in the literature according to the authors knowledge.
In particular, theoretical prediction and subsequent agreement with experimental data for blood droplet evaporation and precipitate height profile
has not been reported in the literature.

The current manuscript is structured as follows. Section 2 discusses the materials and methods in several subsections. Blood sample preparation, experimental setup, and dried residue characterization are discussed in sections 2.1, 2.2, and 2.3, respectively.
Section 3.1 introduces the key results as a global overview. Section 3.2 discusses the coordinate system used and sessile droplet evaporation physics.
Section 3.3 discusses the various stages of blood drop evaporation. Section 3.4 discusses the quantitative modeling and the generalized mechanics of blood drop evaporation. Section 3.5 presents the characterization of dried blood residues, discusses the cracking mechanics due to desiccation stresses,
bacterial distribution and the role of bacterial concentration on the precipitate crack patterns. Section 4 concludes the manuscript.
\begin{figure*}
    \centering
    \includegraphics[scale=0.5]{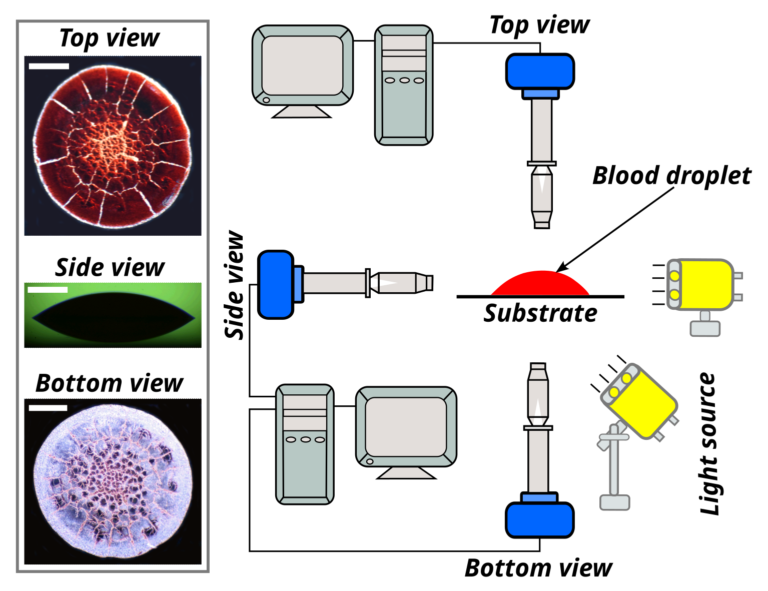}
    \caption{Schematic of the experimental set up and sample top, side and bottom view images of dried blood drop precipitate. Scale bar depicted in white represents $1$ mm.
    }
    \label{Figure1}
\end{figure*}

 \section{Materials and Methods}
 \subsection{Blood sample preparation}
Blood samples were collected from three healthy volunteers (one female and two male) in the age group (25-40 years) with proper consent. 10 ml of whole blood was drawn in K3-EDTA (tripotassium Ethylene Diamine Tetraacetic Acid) vacutainer containers to prevent the blood from clotting. 2-4 ml of the collected blood were sent for several blood tests (CBC (Complete Blood Count), PT (Prothrombin Time), and ESR (Erythrocyte Sedimentation Rate) (refer to supplementary table ST1 for relevant blood report data)) relevant to the mechanics of the various processes that occur during evaporation. Overnight grown stationary phase cultures of \textit{Klebsiella pneumoniae} MH1698 were taken, and their absorbance was measured at OD600nm. Various bacterial concentrations of $10^6$, $10^9$, $10^{12}$ CFU/mL of the bacterial culture were pelleted down at
6000 rpm for 6 minutes. CFU (colony forming units) is a measure of the number number of viable microbial cells in a sample. 
The bacterial concentration parameter was varied in three orders of magnitude to ensure sufficient separation in the concentration parameter space and to avoid overlap due to experimental uncertainty. 
The pellets were washed twice with phosphate-buffered saline (PBS) pH 7.4 and resuspended in EDTA whole blood. For confocal microscopy studies,  \textit{S.} Typhimurium expressing Green Fluorescent Protein (GFP) (pFV-GFP) (STM-GFP) strains were used.
Bacterial concentration larger than $10^{12}$CFU/ml becomes infeasible in our current experiments due to the very large requirement of raw culture of bacterial solution that is diluted repeatedly to attain the desired bacterial concentration. Further, raw bacterial solution concentration higher than $10^{12}$CFU/ml becomes very muddy and viscous which makes handling the samples difficult using typical fluid handling equipments like micro-pipettes and syringes. Viscosity measurements of the samples are performed using cone-and-plate geometry (plate diameter: $40$ mm; cone angle: $1^{\circ}$) of a commerical rheometer (Anton Paar, model MCR302).

\subsection{Experimental setup}
Figure \ref{Figure1} shows a schematic representation of the experimental setup. The experimental setup consists of a blood drop evaporating on a clean glass slide of dimensions $75{\times}25{\times}1$ mm$^3$ (procured from Blue Star). Blood droplet with a volume of $3.4{\pm}0.8{\:}{\mu}$L was placed gently on the glass slide using a micro-pipette. All the evaporation experiments were performed at a relative humidity ($RH$) of $45{\pm}3\%$ and an ambient temperature of $25{\pm}2^{\circ}C$ measured using a TSP-01 sensor, Thorlabs. The complete evaporation process is recorded using color optical diagnostics from three views, i.e., top, side, and bottom. The top and bottom views were operated in reflection mode, whereas shadowgraphy was performed using side-view imaging. Color images of the corresponding views (top, bottom, and side) were recorded using 3 DSLR cameras with an image resolution of 24 megapixels (Nikon D5600) coupled with Navitar zoom lens assembly (2X lens $\times$ 4X adapter tube). A 5W LED light source (Holmarc) and a 50W mercury lamp were used to provide uniform illumination for the three imaging views. A spatial resolution of 1${\mu}$m/pixel was used for the optical color imaging. Consecutive images were captured at a time delay of 10 seconds.
Typical sample top, side, and bottom view images are shown in figure \ref{Figure1} respectively (refer to supplementary movies for temporal data of drop evaporation). The scale bar in white denotes 1mm. Sufficient number (at the minimum of 10 trials are conducted for each experimental parameter) of experimental runs were conducted to ensure a statistical significant dataset.
The instantaneous geometrical parameters and features (shape descriptors) are extracted using image processing techniques applied to the side, top, and bottom view images using open source software ImageJ \citep{schneider2012nih} and python programming language \citep{van2009python}. 

The water content and the delamination height were computed from the reflected light received on the bottom DSLR. The change in color in the third stage of evaporation occurs due to the loss of water and its corresponding delamination height. In general, the delamination height is inversely proportional to the water content in the precipitate. To quantify our measurements in arbitrary intensity units, we map the delamination height to the reciprocal of the water content intensity distribution. 
The white light image were decomposed into its constituents RGB spectrum/modes and the strongest energy component (blue for the bottom) were used. The intensity range of $[0,2^{8}-1]$ were mapped to floating point intensity range $[0,1]$ before further processing. The intensity signals were then thresholded and segmented to remove outlier intensity signals. The entire resolution of the image were used to generate a interpolated intensity signal and the corresponding images were smoothened using appropriate band pass filters. The filtered image were remapped to $[0,2^{8}-1]$ intensity range and the resultant image was passed through the (CLAHE) contrast limited adaptive histogram equalization function. The image output were then calibrated with respect to the non-dimensional time 0.67 where everywhere the height was assigned zero as the corresponding intensity was zero except the outer edge of the droplet where the intensity was non zero at certain places due to delamination. The delamination height and water content was hence measured with respect to a certain temporal instant. It is important to note that the above procedure is a relative method and does not correspond to absolute measurements.

All statistical analysis, data visualization and numerical computation were done using in house codes in python \citep{van2009python}.
\begin{figure*}
    \centering
    \includegraphics[scale=0.5]{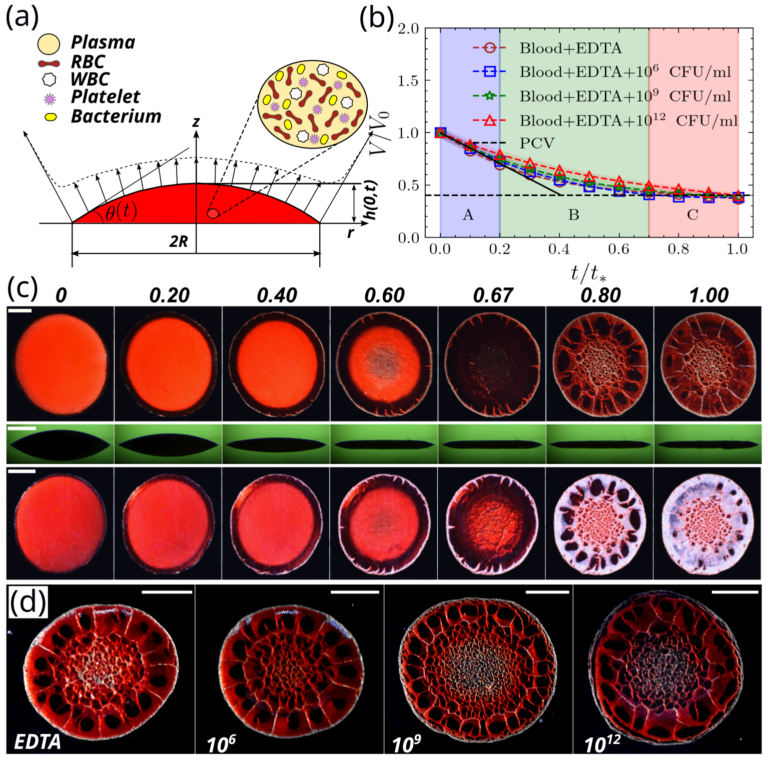}
    \caption{(a) Schematic of the coordinate system, composition of the blood and initial condition of sessile blood drop evaporation.
      (b) Non dimensional volume ($V/V_0$) regression plotted as a function of non dimensional time ($t/t_*$) for whole blood + EDTA, whole blood + EDTA + $10^6$ CFU/ml KP (Klebsiella pneumoniae) bacteria and whole blood + EDTA + $10^9$ CFU/ml KP bacteria respectively. The different stages of sessile blood drop evaporation depicted as A, B and C respectively. The solid black straight line denotes the linear evaporation regime from which the true regression curves deviate at the end of stage A.
      (c) Top view, side view and bottom view time sequence images of the evaporation process respectively. Scale bar for the top, side and bottom view represents $1$ mm  respectively. The timestamps are in non dimensional units ($t/t_*$).
      (d) Top view depicting the final precipitate at $t/t_*=1$ for EDTA, $10^{6}$ CFU/ml, $10^{9}$ CFU/ml, and $10^{12}$ CFU/ml respectively.
      The scale bar denotes $1$ mm respectively.
    }
    \label{Figure2}
  \end{figure*}
\subsection{Dried residue characterization and bacterial distribution}
Scanning Electron Microscopy (SEM) was employed to analyze the surface characteristics of
the deposits, revealing the microcrack patterns through high-resolution imaging. Ultra55 FE-
SEM Karl Zeiss EDS instrument is used. SEM images showed the detailed crack patterns and
stacking of RBCs on the surface. Before the SEM examination, the deposit surfaces were
desiccated for 24 hours and then coated with a thin layer of gold, approximately 10 nm thick,
to facilitate SEM and profilometric studies.
A non-contact optical profilometry method
was utilized to assess the deposit's geometrical features. The measurements were conducted using a Taylor Hobson 3D surface and film
thickness optical profiler. Additionally, the TalySurf CCI, a commercial optical metrology
software, was used to process the data.
The distribution of pathogens within the deposits was also studied using the Andor Dragonfly
confocal microscope system.
The bacteria were labelled with Green Fluorescent Protein
(GFP), which has excitation and emission maxima at 375 nm and 509 nm, respectively.
Fluorescent images were captured using a 10x objective, providing a field of view of $1450{\mu}$m ${\times}$ $1450{\mu}$m. Imaging was performed across multiple z-planes with a step size of $1{\mu}$m.
The final composite image, obtained by super imposing images from several z-planes and
stitching the superimposed images at various locations to reveal the overall bacterial
distribution within the deposit. Live imaging, captured at a rate of one frame per second,
revealed the movement of bacteria within the droplet during the drying process, both near the
droplet edge and at the centre (refer to supplementary movie4 - movie7).
The fluorescence tagged bacteria were tracked to generate the individual trajectory path lines using the TrackMate ImageJ plugin \citep{tinevez2017trackmate,ershov2022trackmate} (refer to supplementary movie8 - movie9).

\section{Results and Discussions}
\subsection{Global overview}
In this work, we study the mechanics of sessile blood drop evaporation (figure \ref{Figure2}) using EDTA-based pure and bacteria-laden whole blood with varying levels of bacterial concentration theoretically and experimentally using lubrication theory, color optical imaging, and micro/nano-characterization.
The concentration ranges from, that are
typically found in living organisms ($c{\leq}10^{9}$ CFU/ml) 
to exceedingly high concentration ($c{\sim}10^{12}$ CFU/ml).
Figure \ref{Figure2}(a) depicts the schematic representation of an evaporating sessile droplet. Figure \ref{Figure2}(b) represents the normalized volume regression of the evaporating droplet for various blood samples (pure and bacteria-laden). We observe that the volume regression is independent of the bacterial concentration range typically found in living organisms 
($c{\leq}10^{9}$ CFU/ml). For exceedingly high bacterial concentration ($c{\sim}10^{12}$ CFU/ml), the evaporation rate reduces slightly, however the changes are not drastic.
The normalized volume asymptotes the packed cell volume (PCV) of the blood sample at the end of evaporation. We further observe from figure \ref{Figure2} (b) that the transient evaporation process can be subdivided into three stages (A, B, and C) based on the evaporation rate. 
The first stage of evaporation  (stage A) ends when the evaporation curves deviate from the initial linear evolution of a pure sessile evaporating droplet with asymptotic form of normalized volume regression $V/V_0{\sim}1-{\gamma}t$, where ${\gamma}$ is the scale for the slope of the regression. The end of second stage (stage B) and the beginning final stage (stage C) occurs when the normalized droplet volume approaches the packed cell volume asymptotically. Stage C represents the final and very slow evaporation rate where processes like major crack formation and precipitate delamination occurs. Figure \ref{Figure2}(c) represents the 
typical top, side, and bottom view image sequences of the evaporation process (refer to supplementary movie1, movie2, and movie3 to view the evaporation process through top, side and bottom imaging respectively). 
Stage A (figure \ref{Figure2}(b,c): $t/t_*=0-0.2$) is the fastest where edge evaporation dominates and leads to the formation of a gelated front propagating radially inwards. Here $t$ denotes the instantaneous time and $t_*$ denotes the total evaporation time scale comprising of all the three stages. The radially outward capillary flow generated by drop evaporation causes the RBCs to accumulate at the outer edge of the droplet. The intermediate stage B (figure \ref{Figure2}(b,c): $t/t_*=0.2-0.7$) consists of gelation of the entire droplet due to the radially inward propagating gelation front and the simultaneous reduction of droplet height $h(r,t)$ and contact angle ${\theta}(t)$ (refer to figure 2(a) for the various shape descriptors of a sessile droplet). The radially inward propagating gelation front leads to RBC deposition, causing the formation of a thick rim around the droplet periphery known as corona \citep{brutin2011pattern}. 
Gelation in stages A and B occurs due to the sol-gel phase transition. The sol-gel phase transition occurs when the solute concentration (RBCs here) becomes larger than a critical concentration. We unearth that the gelation of the entire droplet occurs in stage B. At the end of stage B, the gel formed contains some trace amount of water. The trace amount of water makes the gel wet and stick (laminated) to the glass substrate due to the hydrophilic nature of the glass. Stage C (figure \ref{Figure2}(b,c): $t/t_*=0.7-1.0$) is the final slowest stage of evaporation, where the wet-gel formed during stage A and stage B transforms into dry-gel due to a very slow evaporation process. Due to the continued loss of water from the wet-gel, the laminated regions adhered to the glass substrates undergo delamination. On further desiccation, the drying droplet results in high azimuthal stress, forming radial cracks in the corona region. Mudflat cracks are observed in the center part of the evaporating droplet, where the drop thickness is relatively small and curvature is negligible. We further show that the drop evaporation rate and the corresponding dried residue pattern do not change appreciably within the parameter variation of the bacterial concentration typically found in living organisms (figure \ref{Figure2}(b)).
However, at exceedingly high concentration of $10^{12}$ CFU/ml, the crack pattern in the peripheral corona region deviates from the patterns found at relatively lower concentration ($c{\leq}10^9$ CFU/ml). Figure 2(d) shows the top view of the final precipitate at $t/t_*=1$ for EDTA, $10^6$ CFU/ml, $10^9$ CFU/ml and $10^{12}$ CFU/ml respectively.
From figure 2(d) it clearly evident that the cracks in the corona region of the dried precipitate deviates from radial direction at very high bacterial concentration.

\subsection{Coordinate system and sessile droplet evaporation physics}
Figure \ref{Figure2}(a) shows the schematic of the evaporating droplet and the associated coordinate system. An axisymmetric cylindrical coordinate system ($r-z$) is used to describe and analyze the evaporation process quantitatively. Here, $r$ represents the radial coordinate, and $z$ represents the axial coordinate, the vertical axis along which the droplet liquid-vapor interface profile would be symmetric. In general the sessile droplet can be described by certain geometrical shape descriptors like contact angle (${\theta}$), contact radius ($R$), and drop central height ($h(0,t)$) for a particular initial droplet volume and substrate.
For small Bond number ($Bo={\Delta}{\rho}gl^2/{\sigma}<1$), the droplet geometry can be approximated by a spherical cap geometry \citep{hu2002evaporation,gennes2004capillarity,clift2005bubbles}, where ${\Delta}{\rho}$ is the density difference at the liquid air interface, $g$ is the acceleration due to gravity, $l$ is the characteristic length scale, and ${\sigma}$ is the liquid air surface tension. For the current experiments, using ${\Delta}{\rho}{\sim}\mathcal{O}(10^3)$ kg/m$^3$, ${g}{\sim}\mathcal{O}(10)$ m/s$^2$, ${l}{\sim}R{\sim}\mathcal{O}(10^{-3})$ m, and ${\sigma}{\sim}\mathcal{O}(7{\times}10^{-2})$ N/m, we have $Bo{\sim}\mathcal{O}(10^{-1})<1$ and hence spherical cap assumption can be used to describe the evaporating droplet. The interface profile of the sessile droplet is therefore given by \citep{hu2002evaporation}
\begin{equation}
    h(r,t)=\sqrt{\frac{R^2}{{\sin}^2{\theta}} - r^2} - \frac{R}{{\tan}{\theta}}
\end{equation}
and the corresponding drop volume is given by
\begin{equation}
    V(t) = {\pi}h(0,t)[3R^2 + h^2(0,t)]/6    
\end{equation}
where $h(0,t)=R{\tan}({\theta}(t)/2)$.
In case of blood drop evaporation, the spherical cap assumption would be valid upto a particular time scale ($t/t_*{\sim}0.5$) after which deviation from spherical cap geometry becomes too large due to the effect of gelation. Initially (i.e, $t/t_*=0$), the evaporative flux profile at the drop interface is highest at the drop contact line and least at the drop center (refer to figure 2(a)). Diffusion-limited evaporation model can be used to approximate isothermal sessile drop evaporation for droplets in the size range of mm (i.e., $R{\sim}\mathcal{O}(10^{-3})$m) to a very high degree of accuracy. The size where molecular reaction kinetics dominates diffusion evaporation phenomena occurs approximately at length scales of the order of $100$nm \citep{10.1063/5.0196219}, which is considerably very small with respect to the typical length scale that we have in our current experiments, therefore conforming to diffusion-limited drop evaporation. Drop evaporation, in general, can occur in various modes like CCR (constant contact radius) and CCA (constant contact angle) \citep{wilson2023evaporation} depending on the wettability of the substrate. For hydrophilic substrates with an initial acute contact angle, as is the present case, drop evaporation generally occurs in the CCR mode of evaporation. In the CCR mode of evaporation, the droplet's three-phase contact line is pinned throughout the evaporation process while the contact angle decreases with time. For the present case of evaporating whole blood droplets, the evaporation occurs in CCR mode, as can be observed from the image snapshots in figure \ref{Figure2}(c). The evaporation in CCR mode generates an capillary flow inside the droplet. For \textit{thin} (i.e. small contact angles) droplets the conservation of mass for the solvent in the evaporating droplet (water in general) is given by \citep{wilson2023evaporation}
\begin{equation}
    \frac{{\partial}h}{{\partial}t}  + \frac{1}{r}\frac{{\partial}(rh<u>)}{{\partial}r} = - \frac{J}{\rho}
\end{equation}
where $<u>$ is the height average radial velocity, ${\rho}$ is the liquid density and the evaporation flux $J$ is given by
\begin{equation}
    J = \frac{2Dc_v(1-RH)}{{\pi}\sqrt{R^2-r^2}}
\end{equation}
for $0{\leq}r<R$, where $D$ is the diffusivity of water vapor in air, $c_v$ is the saturation concentration of water vapor at the droplet interface at a particular temperature, and $RH$ is the relative humidity. Using equation (3.4) in equation (3.3), the height averaged radial velocity can be computed as \citep{wilson2023evaporation}
\begin{equation}
    <u> = \frac{4Dc_v(1-RH)}{{\pi}{\rho}{\theta}r}\left[1-\left(1-\frac{r^2}{R^2}\right)^{3/2}\right]\left(1-\frac{r^2}{R^2}\right)^{-1/2}
\end{equation}
where $(r,{\:}{\theta}){\neq}0$. For particle laden droplets the radially outward capillary flow can cause edge deposition like the coffee ring effect \citep{deegan1997capillary}. For dilute concentration of particles and assuming negligible particle diffusion the particle concentration $c$ would satisfy \citep{d2023effect,wilson2023evaporation}
\begin{equation}
    \frac{{\partial}c}{{\partial}t}  + {<u>}\frac{{\partial}c}{{\partial}r} =  \frac{Jc}{{\rho}{h}}
\end{equation}
The assumption of negligible particle diffusion could be understood from the Peclet number of the RBCs. Peclet number in transport processes is the ratio of advective transport rate to diffusive transport. The major cellular component responsible for pattern formation is the aggregation and transport dynamics of the red blood cells (RBCs). Majorly the RBCs are transported across the droplet through capillary driven advection and diffusion of RBCs are negligible. The essential transport mode could be understood by analyzing the scales of Peclet number for RBCs. The Peclet number for RBCs is defined as $Pe{=}uL/D_{RBCs}$, where $u$ is the characteristic velocity scale, $L$ is the characteristic length scale of the flow and $D_{RBCs}$ is the diffusivity of the RBCs in blood plasma. The scale of the diffusivity $D_{RBCs}$ could be calculated using the Stokes-Einstein equation as $D_{RBCs}{\sim}K_BT/6{\pi}{\eta}_Pr_{RBCs}$ \citep{sadhal2012transport}, where $k_B$ is the Boltzmann constant, $T$ is the absolute temperature in Kelvin, ${\eta}_P$ is the plasma viscosity, $r_{RBCs}$ is the characteristic length scale of a RBC. Using $k_B{\sim}1.38{\times}10^{-23}$ J/K, $T{\sim}300$ K, ${\eta}_P{\sim}1.3{\times}10^{-3}$ Pa s, and $r_{RBCs}{\sim}4{\times}10^{-6}$ m, the diffusivity becomes $D_{RBCs}{\sim}4.22{\times}10^{-14}$. Using $u{\sim}\mathcal{O}{(10^{-6})}$ m/s, and $L{\sim}\mathcal{O}{(10^{-3})}$ m, the corresponding Peclet number evaluates to $Pe{\sim}{2.37}{\times}10^4$ indicating negligible effect of diffusion compared to advection.
Diffusion can be important in a very thin region near the pinned contact line as was shown in the latest works of Moore et al. \citep{moore2021nascent,moore2022nascent,moore2023gravitational}. Moore et al. investigated the nascent coffee ring which forms when solute diffusion counters and balances advection. For very small Capillary number and large Peclet number (as is the case for our current experiment), the importance of solute diffusion is confined to a very thin region (boundary layer) near the pinned contact line. Also, as can be observed through the non-dimensional advection-diffusion equation, the Peclet number appears as a coefficient in the denominator of the diffusion term and can be neglected for very high Peclet number in general. Further, diffusion being a physical process governed by random motion of particles, the particle size is an important parameter. Diffusion is important for small particle sizes typically of the order of ${\sim}\mathcal{O}(10-10^2)$ nm. In the current experiment, the typical length scale of the major particles (RBCs) are of the order of ${\sim}\mathcal{O}(10){\:}{\mu}$m, which is at least $100$ bigger than the particles for which diffusion becomes important. Therefore, we can safely neglect the diffusion of RBCs in the evaporating droplet.

Using equation (3.4), (3.5) and the method of characteristics for solving the first order PDE for $c$ we have
\begin{equation}
    c = c_0\left[\left(\frac{{\theta}}{{\theta}_0}\right)^{-3/4} + \left(1 - \frac{r^2}{R^2}\right)^{3/2} - 1 \right]^{1/3}\left(1 - \frac{r^2}{R^2}\right)^{-1/2}
\end{equation}
where $c_0=c_0(r)$ is the initial radial concentration field along the droplet radius. It is important to note that the concentration computed from equation (3.7) is only valid for \textit{thin} droplets and dilute solution. The variation of evaporation flux, depth average radial velocity and depth average particle concentration as a function of radial coordinate is shown in figure S1-S3 of the supplementary material respectively. It is important to note that the radial variation of evaporation flux, velocity and particle concentration from equation (3.4, 3.5, and 3.7) is a monotonically increasing function. Further from equation (3.5, 3.7) we can observe the singularity at the contact line $r=R$ for both $<u>$ and $c$. For thin drops and dilute concentration the evaporative flux is independent of the height profile and the particle concentration, unlike blood drop evaporation where the dependence exists. For blood droplet evaporation the solution for $J$, $h$ and $c$ can not be obtained independently as depicted above and has to be obtained in a coupled manner (refer to section 3.4 for the details). This is due to the non-monotonic nature of the evaporation flux due to gelation. Further the non-trivial coupling between $c$, $J$ and $h$ also plays an important role as discussed in later portion of the text (refer to section 3.4).
From figure \ref{Figure2}(b), we can observe that the entire CCR mode of the evaporation process can be further subdivided into three stages, A, B, and C, based on the descending rate of evaporation.
Figure \ref{Figure2}(b) represents the graph of normalized volume ratio ($V/V_0$) with normalized time ($t/t_*$) for the evaporating drop. Here $V$ denotes the instantaneous drop volume, and $V_0$ denotes the initial drop volume. Here, $t$ denotes the instantaneous time, and $t_*$ denotes the complete evaporation time scale (defined experimentally as the time instant where no significant changes are observed in drop morphologies in all three views).
The end of stage A (beginning of stage B) occurs when the regression curves deviate from the intial asymptotic linear evaporation regime of a pure sessile droplet. The initial linear regime is determined by curve fitting the initial regression data for all bacterial concentration and is shown
as a black solid line in figure 2(b). The goodness of fit characterized by the coeffecient of determination is $R^2=0.95$. Stage B ends (stage C begins) when $V/V_0$ approaches the packed cell volume (PCV).
From figure \ref{Figure2}(b,c), we can observe that stage A corresponds to $t/t_*=0-0.2$, stage B corresponds to $t/t_*=0.2-0.7$ and stage C corresponds to $t/t_*=0.7-1.0$. The mechanics of the individual stages during drop evaporation are discussed below.
\begin{figure*}
    \centering
    \includegraphics[scale=0.45]{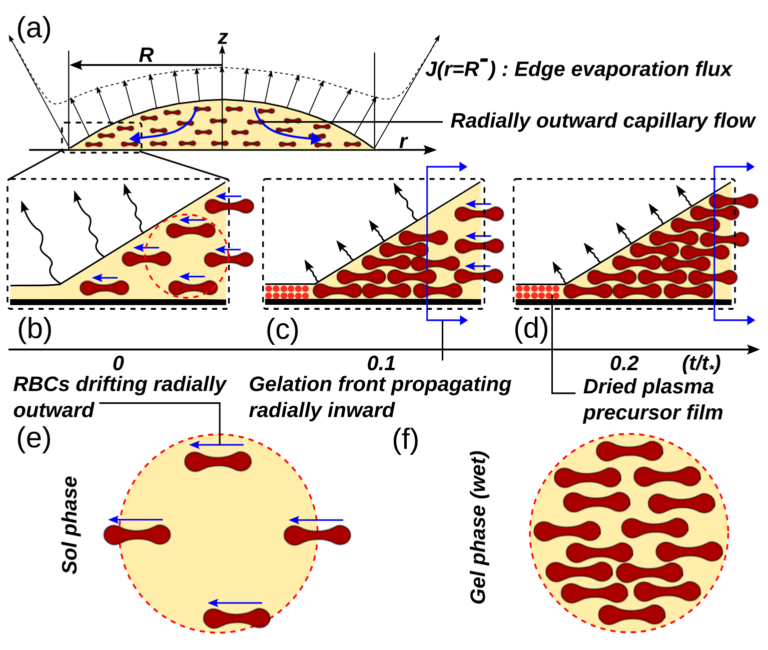}
    \caption{Schematic representation of the various processes occurring in stage A of blood droplet evaporation. (a) Initial configuration of the sessile blood droplet at $t/t^*=0$ depicting the evaporative flux and the radial outward capillary flow inside the evaporating droplet. (b,c,d) Magnified view of the outer edge of the droplet depicting the precursor film and the outward transport of RBCs towards the edge at (b) $t/t^*=0$, (c) $t/t^*=0.1$, and (d) $t/t^*=0.2$ respectively. The blue vertical line shows the gelation front propagating radially inwards. (e) A small control volume (CV) near the three phase contact line depicting the sol phase in which RBCs are present inside the CV and getting transported across the surface of the CV. (f) Wet gel phase in the CV as RBCs concentration increases.}
    \label{Figure3}
\end{figure*}

\begin{figure*}
    \centering
    \includegraphics[scale=0.45]{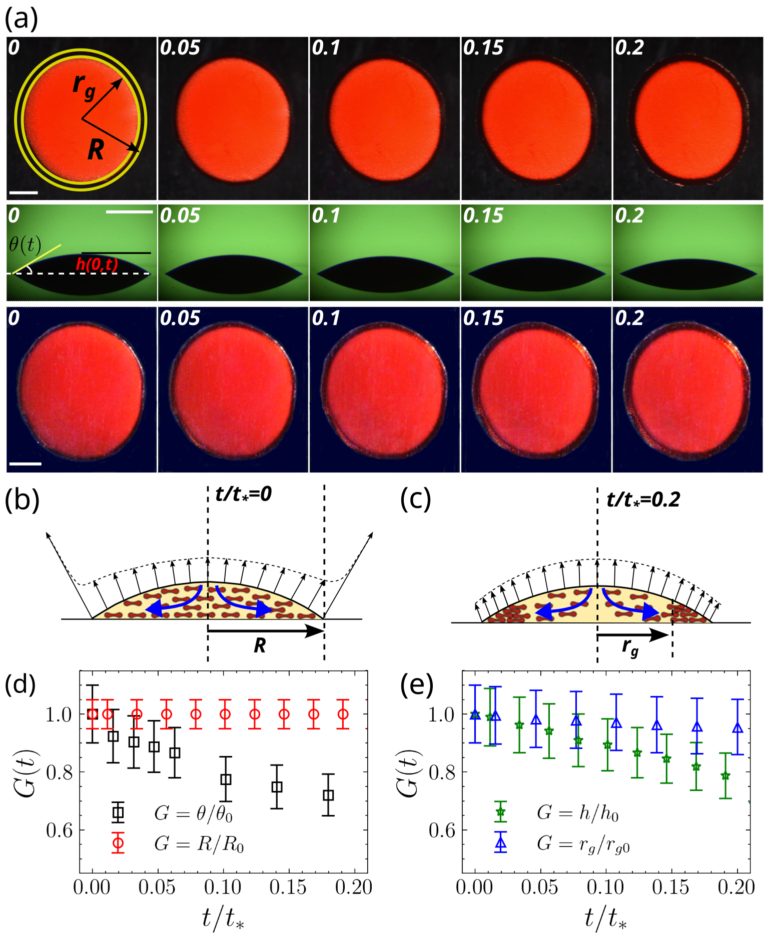}
    \caption{
    (a) Top, side and bottom view time sequence images of stage A at different non dimensional time instants $t/t_*=0,{\:}0.05,{\:}0.1,{\:}0.15,{\:}0.2$ respectively. Scale bar for top, side and bottom view represents 1.2 mm, 1.3 mm and 1 mm respectively. 
    (b) Schematic representing the initial configuration of the evaporating droplet ($t/t_*=0$). (c) Schematic representing the evaporating droplet at the end of evaporation stage A ($t/t_*=0.2$)
    (d,e) Non dimensional geometrical drop parameters $G(t)$ (normalized contact angle (${\theta}/{\theta}_0$), normalized contact radius ($R/R_0$), normalized drop height ($h/h_0$), and normalized gelation radius ($r_g/r_{g0}$)) evolution as a function of non dimensional time $t/t_*$.
    }
    \label{Figure4}
\end{figure*}

\subsection{Stages of blood drop evaporation}
\subsubsection{Stage A}
Figure \ref{Figure3} shows the schematic representation of various processes that occur during stage A. Figure \ref{Figure2}(a) and \ref{Figure3}(a) show the initial configuration of the evaporating blood droplet. The sample blood droplet is a suspension of cellular elements like RBCs, WBCs, platelets, and bacteria (for bacteria-laden drops) suspended in plasma. However, due to the very high number density of RBCs, blood can be approximated as a binary suspension to first approximation. Blood droplet evaporation on hydrophilic substrates occurs in CCR mode with a constant contact radius $R$ due to the pinning of the contact line (refer to side view panel of figure 2(c), 4(a), 5(a), 6(b), and 7(a) and video data (Movie2)). For small drops, the droplet shape at the initial time can be approximated by a spherical cap. The evaporation flux along the droplet interface monotonically increases with radial coordinate $r$ and peaks at the contact line at $t/t_*=0$. The non-uniform evaporation flux along the drop interface generates a radially outward capillary flow that causes the RBCs to be transported towards the contact line (refer to Movie 1,4,5,6,7,8,9). Figure \ref{Figure3}(b,c,d) shows the magnified view near the contact line at $t/t_*=0,0.1,0.2$ respectively. At the initial time of drop evaporation, a very thin precursor film of plasma exists at the outer edge of the droplet, which slowly solidifies (depicted schematically as red filled circles in figure 3(c) and 3(d)) (refer to Movie 1, 3 for top and bottom view video data respectively). The precursor film is devoid of RBCs as the dimension of the precursor film is smaller than a single RBC length scale. The outer radial capillary flow (shown in the blue arrow) (refer to Movie 6, 7) causes the RBCs to accumulate, forming a gel phase.
In Movie 6 and Movie7, the dark regions which moves radially outwards are the RBCs and the green illuminated objects are the bacteria and their
colonies which also move radially outwards through the gaps in between the RBCs due to the mean capillary flow inside the evaporating droplet.
After the initial gelation of the three-phase contact line, a gelation front radially propagating inward is observed (vertical blue line in \ref{Figure3}(c,d)) (refer to Movie 1, 3). Further, the concentration of RBCs increases at the outer periphery, reducing evaporation flux at the drop interface near the contact line. A small dotted circle representing a control volume (CV) near the contact line is shown in figure \ref{Figure3}(b). Figure \ref{Figure3}(e) shows the zoomed-in view of the CV. The blood suspension inside the CV is initially in the sol phase. However, due to the outward capillary flow, the concentration of RBCs inside the control volume increases, leading to a phase transition that forms a wet gel phase (figure \ref{Figure3}(f)). The evaporative flux over the wet gel region reduces significantly as depicted in the schematic by the reduction of the length of the squigly black arrows (figure \ref{Figure3}(b-d)) depicting the evaporative flux. Sol-gel transition initiates at the three-phase contact line as the concentration approaches a critical gelation concentration $c_g$ 
(refer to section 3.4 for a detailed discussion).

Figure \ref{Figure4}(a) shows the top, side, and bottom image sequence for stage A of the evaporating blood droplet in the top, middle, and bottom panels, respectively. The constant contact pinned radius is denoted by $R$ and the gelation radius by $r_g$ (refer to the yellow circles in the top view image panel of figure \ref{Figure4}(a)). The drop contact angle is represented by ${\theta}$, and the drop center height by $h=h(0,t)$. Figure \ref{Figure4}(b,c) shows the schematic representation of the evaporation stage A at $t/t_*=0$ and $t/t_*=0.2$, respectively. Note the increase in concentration of RBCs and the corresponding reduction of evaporation flux depicted schematically in figure 4(c). Figure \ref{Figure4}(d) shows the evolution of various normalized geometrical parameters $G(t)$ like normalized contact angle (${\theta}/{\theta}_0$) and normalized contact radius ($R/R_0$) confirming CCR mode of drop evaporation. Figure \ref{Figure4}(e) also shows the evolution of other normalized geometrical parameters like normalized droplet center height ($h/h_0$) and normalized gelation radius ($r_g/r_{g0}$) during stage A. It can be observed that the droplet height reduction is faster than that of the gelation front propagation. The faster reduction of central height is one of the important causes of the distinct dried droplet residue profile formed in stage C, which will be discussed later in the text.
\begin{figure*}
    \centering
    \includegraphics[scale=0.5]{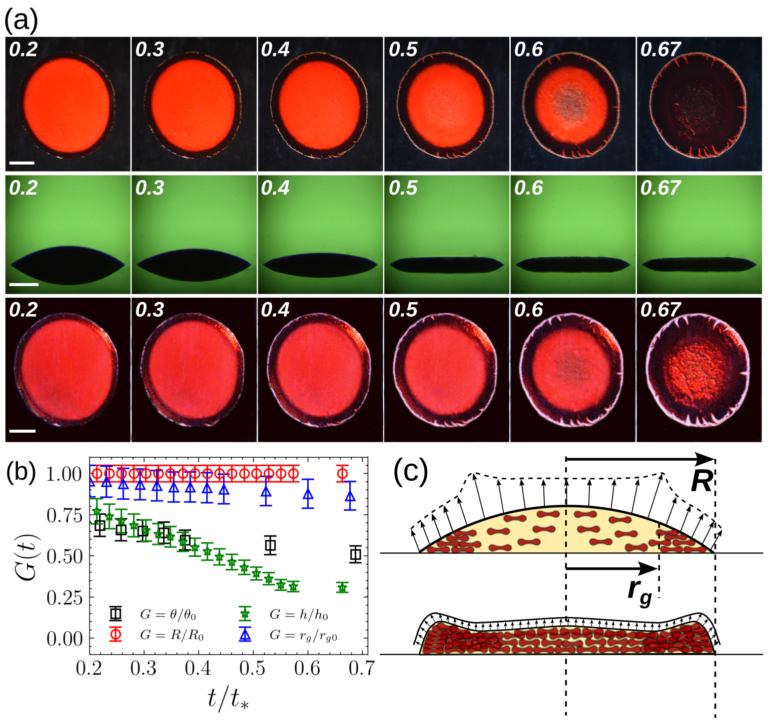}
    \caption{
    (a) Top, side and bottom view time sequence images of stage B at various non dimensional time instants $t/t_*=0.2,{\:}0.3,{\:}0.4,{\:}0.5,{\:}0.6,{\:}0.67$ respectively. The Scale bar represents 1 mm. (b) Non dimensional geometrical drop parameters $G(t)$ (normalized contact angle (${\theta}/{\theta}_0$), normalized drop height (${h}/h_0$), normalized contact radius ($R/R_0$) and normalized gelation radius ${r_g}/{r_{g0}}$) evolution as a function of non dimensional time $t/t_*$. (c) Schematic representing the evaporating droplet at the beginning of stage B ($t/t_*=0.2$) and end of stage B ($t/t_*{\sim}0.67-0.7$) respectively.
    }
    \label{Figure5}
\end{figure*}

\begin{figure*}
    \centering
    \includegraphics[scale=0.5]{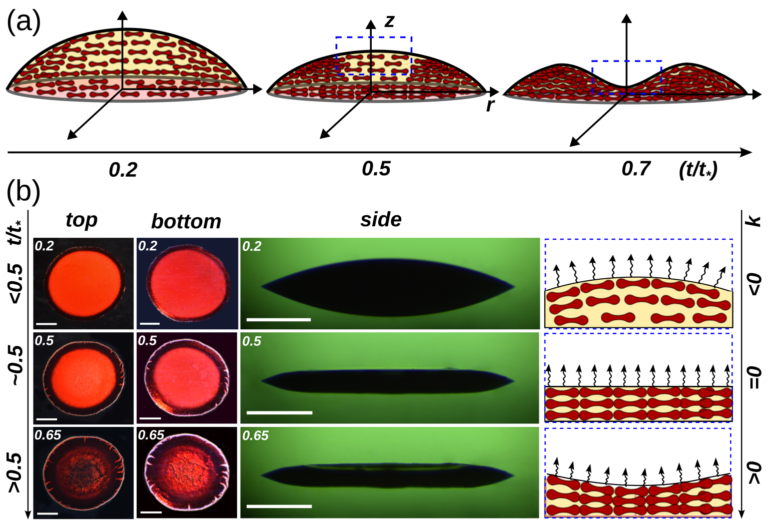}
    \caption{
    (a) 3D Schematic representation of blood droplet evaporation in stage B from $t/t_*=0.2-0.7$.  (b) Top, bottom, side and schematic image sequence depicting curvature evolution and precipitate formation at $t/t_*{<}0.5,{\:}{\sim}0.5,{\:}{>}0.5$ respectively. The scale bar represents 1 mm respectively.
    }
    \label{Figure6}
\end{figure*}
\subsubsection{Stage B}
Stage B of the evaporation process spans from $t/t_*=0.2-0.7$ and has the longest duration. In this stage, the gelation front propagates further radially inwards. As the droplet evaporates in CCR mode, the contact line is pinned, but the droplet contact angle and its corresponding maximum height decreases monotonically. As the droplet evaporates, the outward capillary flow causes a depletion of RBCs in the center, allowing the droplet interface to change its curvature smoothly. 
Figure \ref{Figure5}(a) shows the top, side, and bottom view image sequence during stage B of evaporation. The timestamps are normalized time ($t/t_*$). Figure 5(b) shows the droplet nondimensional geometrical parameters $G(t)$ evolution as a function of normalized time. Various parameters like normalized contact angle (${\theta}/{\theta}_0$), normalized height $h/h_0$, normalized contact radius ($R/R_0$), and normalized gelation radius ($r_g/r_{g0}$) is plotted in black, green, red, and blue respectively. It is important to note that at approximately ($t/t_*{\sim}0.5-0.6$) there is a significant reduction of the droplet height compared to its contact angle signifying the deviation of the droplet shape from spherical cap approximation. Figure 5(c) shows a schematic representation of the starting ($t/t_*=0.2$) and ending state ($t/t_*{\sim}0.67-0.7$) of stage B, respectively. $R$ denotes the contact radius and $r_g$ denotes the gelation front radius. Notably, the distinct color change in the top view at the end of stage B ($t/t_*=0.67$) signifies the completion of the gelation of the entire droplet. At the end of stage B, the entire droplet undergoes a sol-gel phase transition to a final wet gel form. The wet gel phase contains very small (trace) amounts of liquid water that evaporates in the next stage, C. We can also observe from the side view image panel that there is negligible change in droplet geometry, signifying the completion of gelation for the entire droplet.

Figure \ref{Figure6} shows a schematic representation of the various processes occurring during stage B of evaporation. Figure 6(a) shows a 3D schematic of the vertical cross section of the droplet interface curvature change as
evaporation proceeds through stage B at $t/t_*=0.3,0.5,0.7$. Figure 6(b) shows the top, bottom, side and the schematic representation of the droplet interface undergoing monotonic curvature ($k$) change from $k<0$ to $k=0$, to $k>0$ at $t/t_*<0.5$, $t/t_*{\sim}0.5-0.6$, and $t/t_*>0.6$ respectively.

\begin{figure*}  
\includegraphics[scale=0.45]{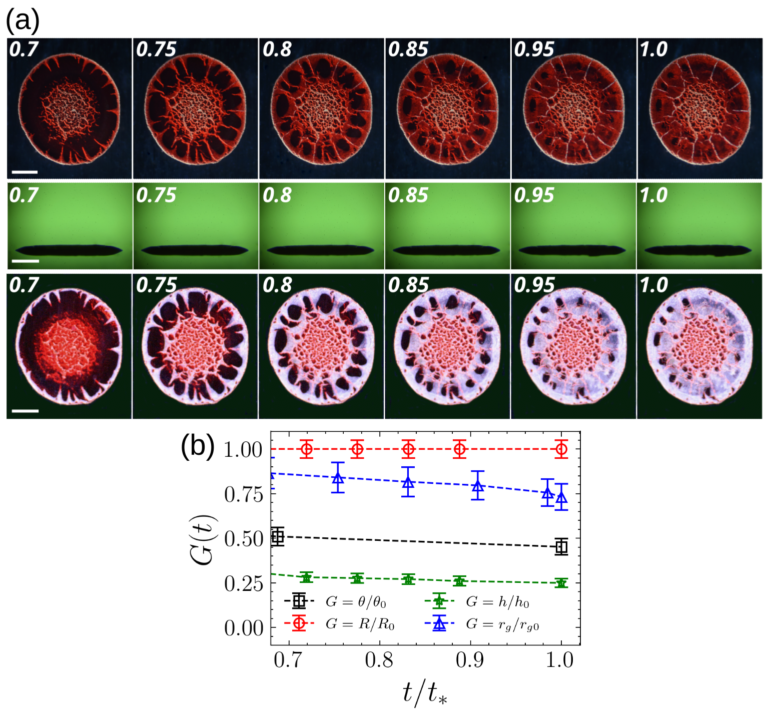}
\centering
\caption{
(a) Top, side and bottom view time sequence images of stage C at non dimensional time instants $t/t_*=0.7,{\:}0.75,{\:}0.8,{\:}0.85, {\:}0.95,{\:}1.0$ respectively. The scale bar represents 1 mm. (b) Non dimensional geometrical drop parameters $G(t)$ (normalized contact angle (${\theta}/{\theta}_0$), normalized drop height (${h}/h_0$), normalized contact radius ($R/R_0$) and normalized gelation radius ${r_g}/{r_{g0}}$) evolution as a function of non dimensional time $t/t_*$.
}
\label{Figure7}
\end{figure*}

\begin{figure*}  
\includegraphics[scale=0.45]{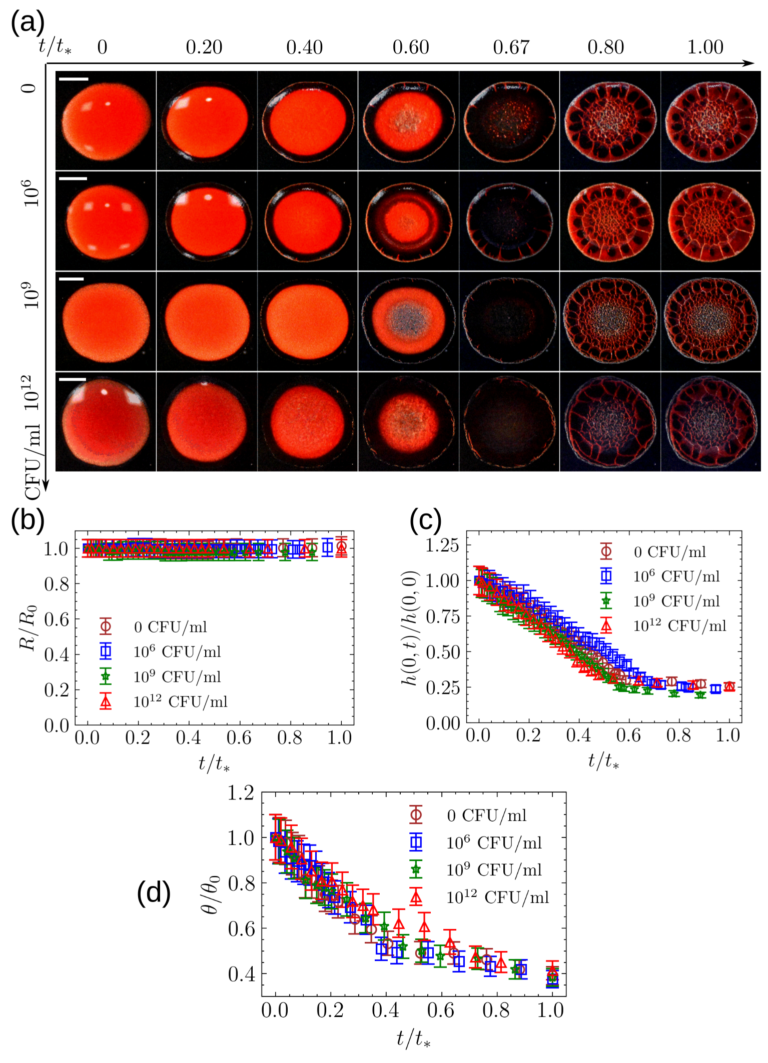}
\centering
\caption{
(a) Comparative time series depicting top view of evaporating blood droplet for $0$, $10^6$, $10^9$ and $10^{12}$ CFU/ml for $t/t_*=0,{\:}0.20,{\:}0.40,{\:}0.60,{\:}0.67,{\:}0.80,{\:}1.00$ respectively. 
(b) Normalized droplet contact radius as a function of normalized time for $0$, $10^6$, $10^9$ and $10^{12}$ CFU/ml bacterial concentration respectively.
(c) Normalized droplet central height as a function of normalized time for $0$, $10^6$, $10^9$ and $10^{12}$ CFU/ml bacterial concentration respectively.
(d) Normalized droplet contact angle as a function of normalized time for $0$, $10^6$, $10^9$ and $10^{12}$ CFU/ml bacterial concentration respectively.
}
\label{Figure8}
\end{figure*}

\begin{figure*}
\includegraphics[scale=0.45]{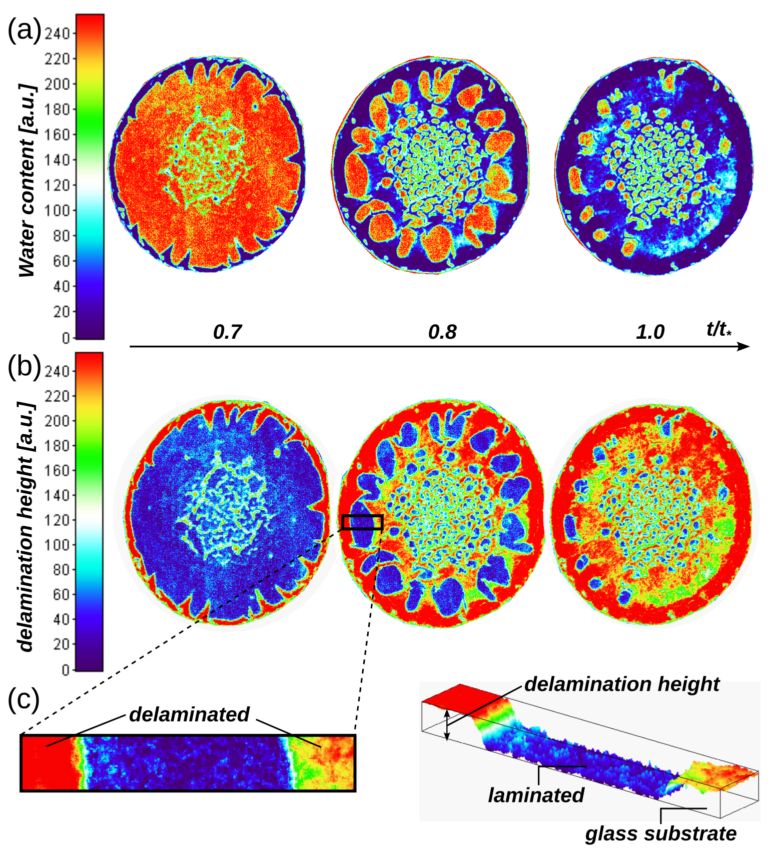}
\centering
\caption{
(a) Evolution of entrapped water evaporation as a function of non-dimensional time in stage C represented as an intensity color map. (b) Evolution of the 2D delamination height as a function of non-dimensional time in stage C represented as an intensity color map. (c) Magnified view depicting a typical laminated and delaminated region at the corona of dried blood residue at $t/t_*=0.8$.
}
\label{Figure9}
\end{figure*}

\subsubsection{Stage C}
Figure \ref{Figure7}(a) shows the image sequence of the top, side, and bottom view of the desiccating drop, respectively. Notice the distinct color change in the top and bottom view as the drop transforms from wet gel to dry gel.
The color change is the signature for the loss of water (and corresponding delamination from the substrate) from the wet gel precipitate and transforming it into a dry gel form.
From both the top and bottom view, we can observe the evaporation process of the trace amounts of water present at the beginning of stage C. Dessication stresses are developed as the drop undergoes further evaporation, forming various types of cracks. Radial cracks are observed in the thicker outer rim region, whereas mudflat cracks are observed in the center and the contact line region. \ref{Figure7}(b) shows the slow variation of the nondimensional geometric parameters such as contact angle, drop center height, contact radius, and gelation radius as a function of nondimensional time. It is important to contrast the slow variation of the geometrical parameters in stage C from the relatively fast variation in stages A and B.
The role of bacteria on blood droplet evaporation and final precipitate can be understood by comparing the time series images of the evaporating droplet. Figure 8(a) shows the image snapshots for $0$, $10^6$, $10^9$, and $10^{12}$ CFU/ml bacterial concentration at non dimensional time $t/t_*=0,{\:}0.20,{\:}0.40,{\:}0.60,{\:}0.67,{\:}0.80,{\:}1.00$. Figure 8(b), 8(c), and 8(d) depicts the time evolution of normalized contact radius $R/R_0$, normalized central height $h(0,t)/h(0,0)$, and normalized contact angle ${\theta}/{\theta}_0$ for the evaporating droplet of $0$, $10^6$, $10^9$, and $10^{12}$ CFU/ml bacterial concentration respectively. From figure 8(b) it is evident that the entire evaporation of the droplet occurs in pinned mode. Using figure 8(b), 8(c), 8(d), and figure 2(b) it is evident that evaporation is independent of bacterial concentration upto $10^9$ CFU/ml. The volume regression curve for $10^{12}$ CFU/ml (figure 2(b)) shows very small decrease in evaporation rate compared to lower concentration upto $10^9$ CFU/ml, however the changes are very small. Although the evaporation characteristics does not change drastically for $10^{12}$ CFU/ml in comparison to $10^9$ CFU/ml, the final crack precipitate pattern of $10^{12}$ CFU/ml changes from that of lower bacterial concentration of $10^9$ CFU/ml (refer to the later section 3.5 and figure 13).    
Figure 9(a) shows the evaporation of the entrapped water in the wet gel phase as a function of nondimensional time represented as an intensity color map in arbitrary units (a.u.) for effective visualization.
Figure 9(b) denotes the time sequence of 2D delamination height represented as an intensity color map in arbitrary units (a.u.) during wet gel to dry gel transformation in stage C. Figure 9(c) depicts a magnified view of the corona region of the dried blood precipitate highlighting regions like laminated and delaminated sections over the glass substrate. The dried residue undergoes a lamination-to-delamination transition as the wet gel transforms into a dry gel. The trace amounts of water present in the wet gel phase (red region in the intensity color map shown in figure 9(a) cause the wet gel to adhere to the glass substrate (blue region in the intensity color map shown in figure 9(b) and 9(c)) due to the hydrophilic nature of glass substrate. However, on further desiccation, the water content in the laminated region decreases and leads to delamination from the substrate due to the phobic nature of the residual glycoproteins found on the surface of the RBCs \citep{brutin2011pattern}. The primary functionality of the glycoproteins is to minimize wettability from a surface.

\subsection{Generalized Mechanics of Blood drop evaporation}
The initial capillary flow initiated due to the non uniform evaporation flux over the drop surface can be modelled using Stokes flow in cylindrical coordinates (refer to figure 2(a) and 3(a)). 
We follow the analysis of Brutin et al., Tarasevish et al., and Sobac et al. \citep{brutin2011pattern,tarasevich2011desiccating, sobac2011structural} very closely for computing the dynamics of drop evaporation.
From the radial component of the Stokes equation along with the thin film approximation (lubrication approximation) we have
\begin{equation}
    \frac{{\partial}p}{{\partial}r} = {\eta}\frac{{\partial}^2u}{{\partial}z^2}   
\end{equation}
In the above equation (3.8) the momentum diffusion is assumed to be negligible in the flow radial direction $r$ in comparison to the vertical $z$ direction due to the aspect ratio of the geometry (thin film approximation). In our experiments we are dealing with droplets whose width is at least one order larger than central height of the droplet and have relatively small contact angles due to hydrophilic substrates. Hence, it is evident from mass conservation that radial velocity will be significantly higher than the vertical component of velocity. The velocity profile in the vertical direction is due to the effect of momentum diffusivity, however the variation in droplet velocity in the radial direction is majorly driven the by the droplet shape (spherical cap). Hence, the effect of radial momentum diffusion will become important for droplets with aspect ratio close to unity and have relatively higher contact angles (hydrophobic substrates) which is unlike the current scenario which justifies the negligible effect of radial momentum diffusion. Therefore,
from the axial component of the Stokes equation we have
\begin{equation}
    \frac{{\partial}p}{{\partial}z}=0
\end{equation}
where $p$ is the pressure, $r$ is the radial coordinate, ${\eta}$ is the drop viscosity, ${u}$ is the radial velocity, and $z$ is the axial coordinate.
Assuming cylindrical symmetry about the vertical axis, the continuity equation can be written as
\begin{equation}
    \frac{1}{r}\frac{{\partial}(ru)}{{\partial}r}+\frac{{\partial}w}{{\partial}z}=0
\end{equation}
The boundary condition for pressure field on the drop surface $z=h(r,t)$ is given by 
\begin{equation}
    p|_{z=h(r,t)}=-{{\sigma}}k
\end{equation}
where ${\sigma}$ is the surface tension and $k$ is the curvature of the drop interface given by
\begin{equation}
    k = \frac{1}{r}\frac{{\partial}}{{\partial}r}\left(r\frac{{\partial}h}{{\partial}r}\right)
\end{equation}
Further, at the drop interface $z=h(r,t)$ we have 
\begin{equation}
    \frac{{\partial}u}{{\partial}z}\Big|_{z=h(r,t)}=0
\end{equation}
owing to the fact that the radial velocity $u$ has negligible variation in the axial direction $z$. The no slip and no penetration boundary condition at the surface of the substrate $z=0$ ensures
\begin{equation}
    u|_{z=0}=0
\end{equation}
and
\begin{equation}
    w|_{z=0}=0
\end{equation}
Using equation (3.10) - (3.15) the radial velocity $u$ can be written as 
\begin{equation}
    u = - \frac{\sigma}{\eta}\frac{{\partial}}{{\partial}r}\left(\frac{1}{r}\frac{\partial}{{\partial}r}\left(r\frac{{\partial}h}{{\partial}r}\right)\right)\left(\frac{z^2}{2}-hz\right)
\end{equation}
The $z$ component of fluid velocity is given by
\begin{equation}
    w = \frac{\sigma}{r}\frac{\partial}{{\partial}r}\left(\frac{r}{\eta}\frac{\partial}{{\partial}r}\left(\frac{1}{r}\frac{\partial}{{\partial}r}\left(r\frac{{\partial}h}{{\partial}r}\right)\right)\left(\frac{z^3}{6} - h\frac{z^2}{2}\right)\right)
\end{equation}
Integrating equation (3.16) with respect to $z$, the droplet height average radial velocity $<u>$ is given as
\begin{equation}
    <u> = \frac{1}{h}\int_0^h u dz = \frac{h^2}{3}\frac{\sigma}{\eta}\frac{{\partial}}{{\partial}r}\left(\frac{1}{r}\frac{\partial}{{\partial}r}\left(r\frac{{\partial}h}{{\partial}r}\right)\right)
\end{equation}
\begin{figure*}
\includegraphics[scale=0.5]{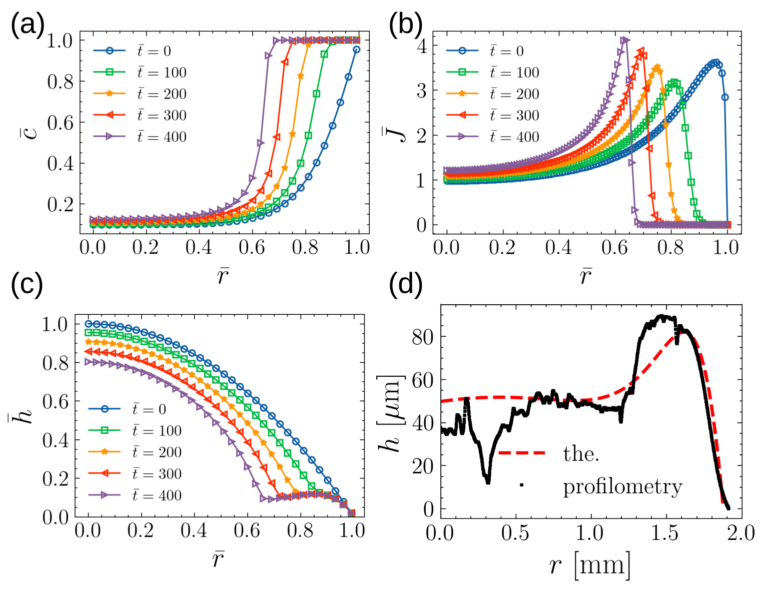}
\centering
\caption{(a) Non dimensional concentration field plotted as a function of non dimensional radial coordinate at various non dimensional time instants $\bar{t}=0,{\:}100,{\:}200,{\:}300,{\:}400$. (b) Non dimensional evaporation flux plotted as a function of non dimensional radial coordinate at various non dimensional time instants $\bar{t}=0,{\:}100,{\:}200,{\:}300,{\:}400$. (c) Non dimensional drop height variation as a function of non dimensional radial coordinate at various non dimensional time instants $\bar{t}=0,{\:}100,{\:}200,{\:}300,{\:}400$. (d) Theoretical and experimental comparison (profilometry) of dried blood droplet precipitate radial height profile.}
\label{Figure10}
\end{figure*}

\begin{figure*}
\includegraphics[scale=0.44]{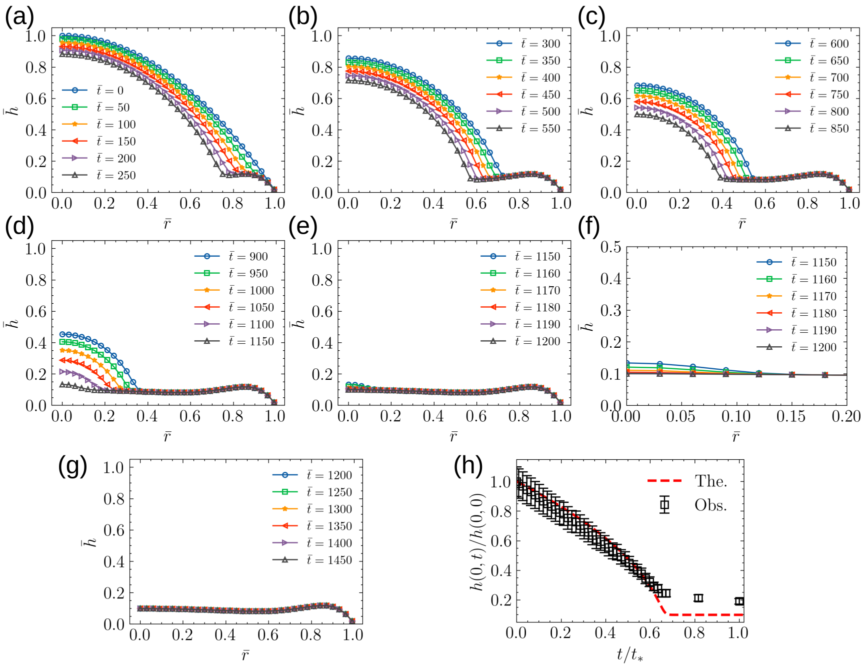}
\centering
\caption{Non dimensional drop height variation as a function of non dimensional radial coordinate at various non dimensional time instants (a) $\bar{t}=0-250$, (b) $\bar{t}=300-550$, (c) $\bar{t}=600-850$, (d) $\bar{t}=900-1150$, (e) $\bar{t}=1150-1200$,  (f) magnified view of non dimensional height profile near $\bar{r}=0-0.2$ for time instants $\bar{t}=1150-1200$, (g) $\bar{t}=1200-1450$. (h) Comparison of theoretical and experimental observation of drop center height variation with time.}
\label{Figure11}
\end{figure*}
Using the continuity equation for the conservation of mass of the volatile component (here solvent majorly is the water content of the blood drop), the average radial velocity from equation (3.18), the dynamical equation for the drop height is given as 
\begin{equation}
    \frac{{\partial}h}{{\partial}t} = - \frac{1}{r}\frac{{\partial}(rh<u>)}{{\partial}r} - \frac{J}{\rho}
\end{equation}
where $J$ is the evaporation flux and ${\rho}$ is the density of the droplet. Similarly, using the continuity equation for the non-volatile component, i.e. solute (conservation of mass for the solute; here the major fraction is the RBCs), the dynamics of the solute concentration $c$ is given as
\begin{equation}
    \frac{{\partial}(hc)}{{\partial}t} = - \frac{1}{r}\frac{\partial}{{\partial}r}(rhc<u>)
\end{equation}
For small drop size where gravity is negligible compared to surface tension effects, the drop shape can be approximated by a spherical cap model. However the spherical cap model leads to singularities in the flow velocity and solute concentration (RBCs here) near the contact line. As a result, we replace the initial drop shape by a paraboloid of revolution without the loss of generality as a initial condition. The initial drop shape is therefore given by
\begin{equation}
    h|_{t=0} = h_f + h_0\left(1 - \left(\frac{r}{R}\right)^2\right)
\end{equation}
where $h_f$ is the drop height at the edge that corresponds to the precursor plasma film. 
\begin{equation}
    h|_{r=R} = h_f
\end{equation}
The sum of $h_f$ and $h_0$ represents the central drop height thickness, i.e., at $r=0$. Owing to the constant contact radius (CCR) mode of drop evaporation, the contact line of the drop is pinned and hence the radial velocity vanishes. The radial velocity also goes to zero at $r=0$ due to vanishing curvature at the center. Therefore the boundary condition for the height average radial velocity $<u>$ to solve equation (3.19) and (3.20) simultaneously is given by 
\begin{equation}
    <u>|_{r=0}= <u>|_{r=R} = 0   
\end{equation}
The symmetry condition for the drop height profile gives
\begin{equation}
    \frac{{\partial}h}{{\partial}r}\Big|_{r=0} = 0    
\end{equation}
Similarly, the symmetry condition for solute concentration is given by
\begin{equation}
    \frac{{\partial}c}{{\partial}r}\Big|_{r=0}=0
\end{equation}
The initial solute concentration field is given by a known function $f(r)$ of radial coordinate
\begin{equation}
    c(r,0) = f(r)
\end{equation}
We solve the equations for the drop height, solute concentration and evaporation flux in non-dimensional coordinate space. The drop height is normalized with respect to initial drop height $h_0$ as $h_f{\ll}h_0$, i.e., $\bar{h}=h/h_0$. The radial coordinate $r$ is normalized with
respect to drop contact radius $R$, i.e., $\bar{r}=r/R$. The height average radial velocity $<u>$ is normalized with respect to the viscous velocity scale $u_c={\eta}_0/{\rho}{h_0}$, i.e. $\bar{u}=<u>/u_c$. Here ${\eta}_0$ is the viscosity of the solvent. The time scale $t$ is normalized by a reference time scale $R/u_c$, i.e., $\bar{t}=tu_c/R$. The vapor flux $J$ is normalized by a reference flux $J_c=k_T{\Delta}T/Lh_0$, i.e., $\bar{J}=J/J_c$; where $k_T$ is the thermal conductivity of the liquid, ${\Delta}T$ is the difference between substrate and the saturation temperature and $L$ is the latent heat of vaporization. The solute concentration $c$ was normalized by the gelation concentration $c_g$, i.e, $\bar{c}=c/c_g$. Using the above normalized variables equation (3.19) and (3.20) can be non-dimensionalized and written as a vector equation with $\bar{F}$ and $\bar{f}$ expressed as a column vector
\begin{equation}
    \frac{\partial{\bar{F}}}{\partial{\bar{t}}}+\bar{u}\frac{\partial{\bar{F}}}{\partial{\bar{r}}} = \bar{f}
\end{equation}
where the column vectors $\bar{F}$ and $\bar{f}$ are given by
\begin{equation}
    \bar{F}=\begin{bmatrix}
        \bar{h}\\
        \bar{c}
    \end{bmatrix}
\end{equation}
and
\begin{equation}
    \bar{f}=\begin{bmatrix}
        -\frac{\bar{h}}{\bar{r}}\frac{{\partial}}{{\partial}\bar{r}}\left(\bar{r}\bar{u}\right) - E\bar{J}\\
        \frac{E\bar{c}\bar{J}}{\bar{h}}
    \end{bmatrix}
\end{equation}
where
\begin{equation}
    \bar{u} = \frac{\bar{h}^2}{3Ca}\frac{{\partial}}{{\partial}\bar{r}}\left(\frac{1}{\bar{r}}\frac{{\partial}}{{\partial}\bar{r}}\left(\bar{r}\frac{{\partial}\bar{h}}{{\partial}\bar{r}}\right)\right)
\end{equation}
Here $E=k_T{\Delta}T/{\epsilon}{\eta}_0L$ and $Ca={\eta}u_c/{\epsilon}^3{\sigma}$ represents the evaporation and capillary number respectively, where ${\epsilon}=h_0/R$. 
The blood viscosity ${\eta}$ dependence on the RBCs haematocrit $Ht$ is given as \citep{lee2007applied,bergel2012cardiovascular}
\begin{equation}
    {\eta} = {\eta}_0\left(\frac{1}{1 - {\alpha}{Ht}}\right)
\end{equation}
where
\begin{equation}
    {\alpha} = 0.076 e ^ {2.49 Ht + (1107/T) e ^{-1.69Ht}}
\end{equation}
and $T$ is the temperature in Kelvin. The ratio of ${\eta}/{\eta}_0$ as a function of $Ht$ is given in figure S4 of the supplementary material (refer to the supplementary). From the supplementary figure S4 it is evident that ${\eta}$ increases monotonically with $Ht$. $Ht$ physically represents the volume fraction of RBCs in the blood droplet.
The blood viscosity measured for the blood samples as a function of bacterial concentration are of the order of $2-4{\times}10^{-1}$ Pa s at a shear rate of $0.1$ s$^{-1}$ which corresponds to a very high viscosity value (refer to the viscosity strain curve obtained from viscosity experiments in supplementary figure S5). 
Therefore as the blood drop evaporates, $Ht$ further increases causing an equivalent increase in blood viscosity and capillary number $Ca$ which further reduces the flow velocity scale $\bar{u}$ inside the evaporating droplet, as $\bar{u}{\propto}Ca^{-1}$. The droplet precipitate thickness therefore becomes very weakly coupled to the flow velocity.
The evaporative mass flux is modeled using techniques from heat transfer analysis \citep{anderson1995spreading} and also assuming that the vapor density vanishes at the sol-gel propagating front; i.e., the evaporative flux goes to zero as the concentration reaches a critical value of $\bar{c}=1$ ($c=C_g$) during sol-gel phase transition at the propagating front. The dimensionless evaporation flux hence can be modeled as
\citep{fischer2002particle,jung2009film,okuzono2010effects,bhardwaj2009pattern}

\begin{equation}
    \bar{J}(\bar{r},\bar{t}) = \frac{1 - \bar{c}^2}{k_d + \bar{h}}
\end{equation}
where $k_d$ is a dimensionless non equilibrium parameter. The range of $k_d$ is such that $k_d{\rightarrow}{\infty}$ for non-volatile liquids and $k_d{\rightarrow}0$ for volatile liquids.
The form of equation (3.33) is based on the observation that higher concentration of RBCs will reduce the evaporation flux and the flux goes to zero at some critical concentration ($\bar{c}=1$) corresponding to a sol-gel phase transition. The squared (quadratic) concentration term is typically not obvious and depends on the requirement of no change in evaporative flux with respect to a concentration change up to a first order change about $\bar{c}=0$. The quadratic term being the lowest order term to satisfy the criterion $\frac{{\partial}\bar{J}}{{\partial}\bar{c}}\Big|_{\bar{c}=0}=0$.
The initial condition for the nondimensional concentration profile required to solve equation (3.27) is given by
\begin{equation}
    \bar{c}(r,0) = 2 - \bar{c}_0 + \frac{2(\bar{c}_0 - 1)}{1 + e^{w(\bar{r}-1)}}
\end{equation}
where $w$ is a measure of a characteristic length over which the concentration of colloidal particles increases rapidly. From the above distribution it is important to note that $\bar{c}(0,0){\simeq}0$ and $\bar{c}(1,0)=1$ suggesting the initial concentration at the edge of the droplet is equal to the gelation concentration (gelation has already occurred at the outer edge of the droplet contact line). Using $D{\sim}2.61{\times}10^{-5}$m$^2$/s, $c_v{\sim}2.32{\times}10^{-2}$kg/m$^3$, $k{\sim}\mathcal{O}(1)$W/mK, ${\Delta}T{\sim}75K$, ${\eta}_0{\sim}5{\times}10^{-3}$Pa s, $L{\sim}2.25{\times}10^6$J/kg, ${\epsilon}{\sim}0.4$, the system of differential equation for $\bar{h}$ and $\bar{c}$ given by equation (3.27) is computed using finite difference methods for $E{\sim}10^{-2}$ and $Ca{\sim}6{\times}10^{-5}$ corresponding to whole blood droplets used in our experiments. 
Figure 10(a) 
shows the non dimensional average concentration profile for various non dimensional time instants $\bar{t}=0,100,200,300,400$. 
Figure 10(b) 
shows the evolution of the evaporation flux profile along the drop interface for various non dimensional time instants $\bar{t}=0,100,200,300,400$. 
The temporal evolution of the average concentration at the outer edge of the droplet depicting sol-gel phase transition (monotonic increase in concentration) is shown in supplementary figure S6. The non-dimensional evaporative flux evolution at the droplet outer edge is also shown in supplementary figure S7 depicting the corresponding monotonic decrease.
Figure 10(c)
shows the non dimensional drop height profile as a function of non dimensional radial coordinate with non dimensional time as a parameter. Figure 10(d) compares the final dried residue droplet thickness profile with the steady state theoretical prediction. The theoretical height profile (red dotted line) conforms with the experimental curve obtained from profilometry (black) within the experimental uncertainty.
Figure 11(a)-(g) shows the non dimensional height profile variation as a function of non dimensional radial coordinate with non dimensional time as a parameter in the range $\bar{t}=0-1450$. 
Figure 11(h) shows the droplet center height variation as a function of non dimensional time $t/t_*$. The theoretical curve in red (labelled as The.) agrees with the experimental observations (labelled as Obs.) within the experimental uncertainty. The variation of the experimental observation from the theoretical prediction after $t/t_*=0.7$ is due to the fact that the experimental drop center height data is extracted from the side view. As a result, after the reduction of the central height to approximately a particular value ($h(0,t)/h(0,0){\sim}0.2$), further reduction in the central height cannot be measured due to the formation of the thick coronal rim around the droplet. The coronal deposit obstructs the view of the center part of the droplet causing the measurement to flatten out much earlier compared to the actual theoretical prediction.

\begin{figure*}
\includegraphics[scale=0.5]{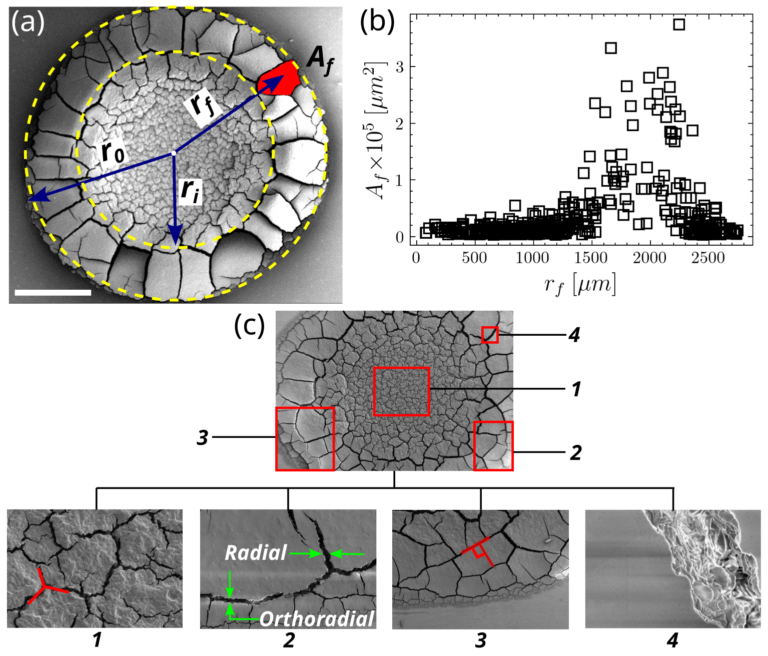}
\centering
\caption{Dried blood precipitate characterization using (a) SEM denoting various kinds of cracks and flakes. Scale bar represents 1.3mm. (b) Radial flake size distribution of dried whole blood drop residue. (c) Cracks characterization in dried whole blood precipitate. The numerals 1-4 represents different viewing region of interest (ROI). 1, 2, 3, 4 represent the center, outer edge of the central region, peripheral region (corona) and a radial crack respectively.}
\label{Figure12}
\end{figure*}

\begin{figure*}
\includegraphics[scale=0.5]{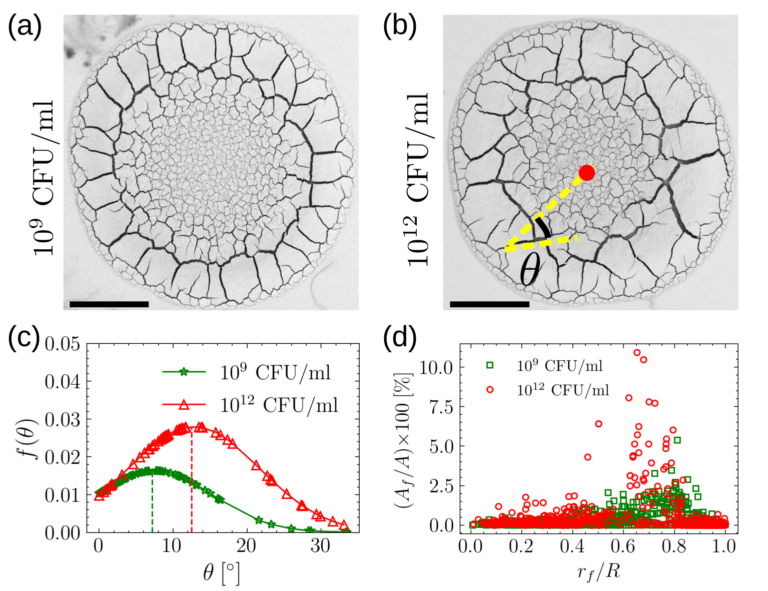}
\centering
\caption{
Comparison of final precipitate pattern using SEM for (a) $10^9$ CFU/ml and (b) $10^{12}$ CFU/ml. The angle ${\theta}$ measures the deviation from a radial crack. (c) The probability density function $f({\theta})$ of the crack angle for $10^9$ CFU/ml and $10^{12}$ CFU/ml. (d) The radial distribution of flake size represented as an area fraction/percentage for $10^9$ CFU/ml and $10^{12}$ CFU/ml.
}
\label{Figure13}
\end{figure*}

\begin{figure*}
\includegraphics[scale=0.5]{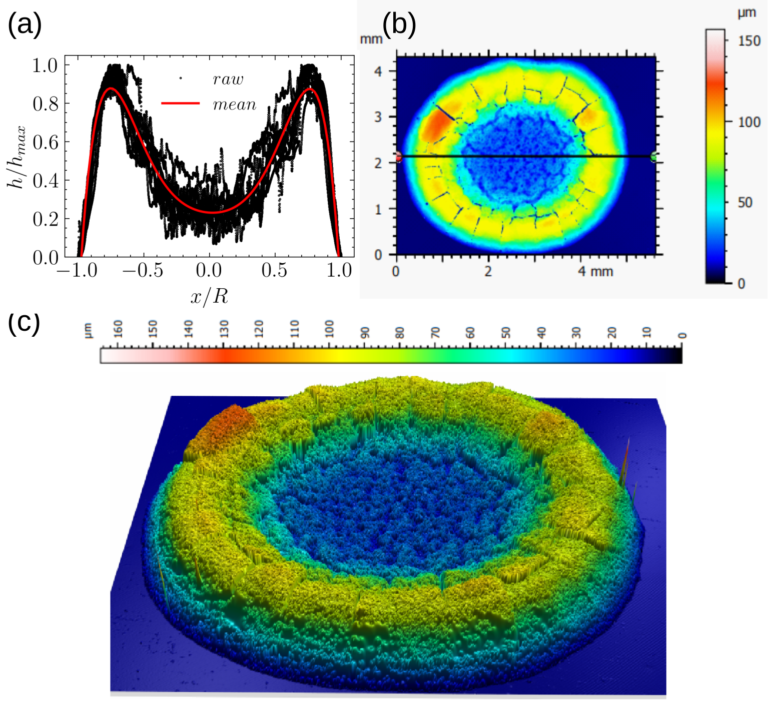}
\centering
\caption{
Dried blood drop surface thickness characterization using optical profilometry.
  (a) 1D surface profile thickness variation along the
  horizontal diametric axis (shown as black line in (b)) of the dried blood droplet precipitate. The black dots represents raw profilometry data as a function of radial coordinate. The red solid curve denotes the mean thickness profile.
  (b) 2D visualization of dried precipitate surface thickness represented as a filled contour plot.
  (c) 3D perspective visualization of the dried residue
  surface and its corresponding thickness.
}
\label{Figure14}
\end{figure*}

\begin{figure*}
\includegraphics[scale=0.50]{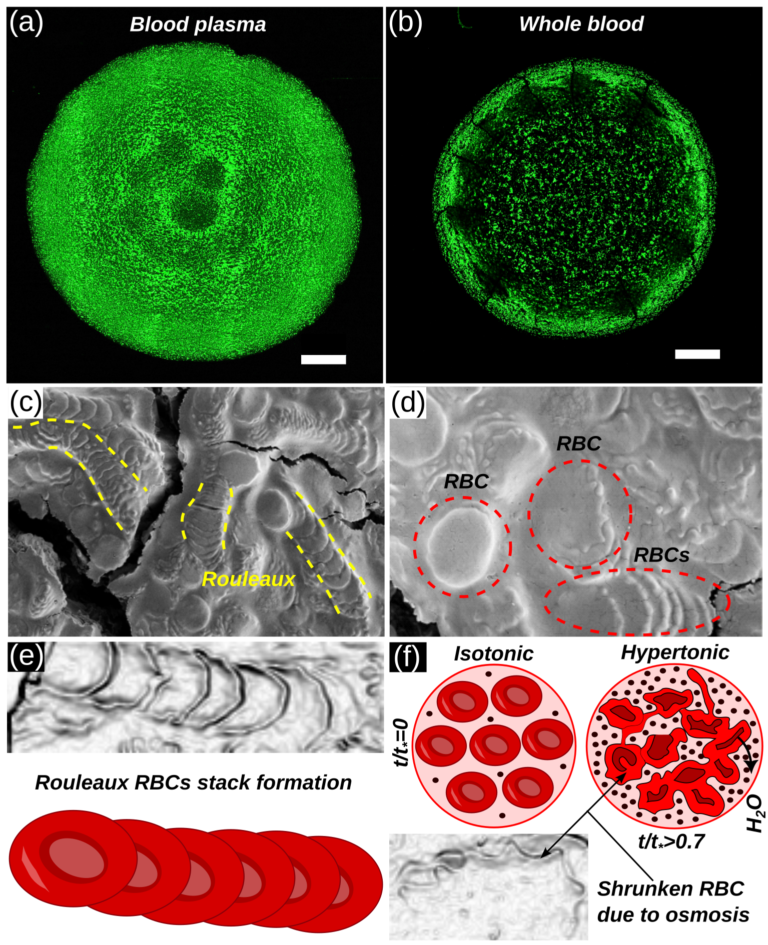}
\centering
\caption{
Confocal fluroscence microscopy images depicting bacterial distribution in the dried precipitate of (a) blood plasma and (b) Whole blood droplet. The white scale bar represents $500{\:}{\mu}$m. (c) SEM image depicting RBCs distribution in the dried blood precipitate. (d) SEM image depicting shrunken RBCs. (e) Schematic representation of Rouleaux stack formation of RBCs. (f) Schematic representing osmosis in RBCs causing shape deformation from biconcave disk to shrunken state at the end state of evaporation (stage C). 
}
\label{Figure15}
\end{figure*}

\begin{figure*}
\includegraphics[scale=0.50]{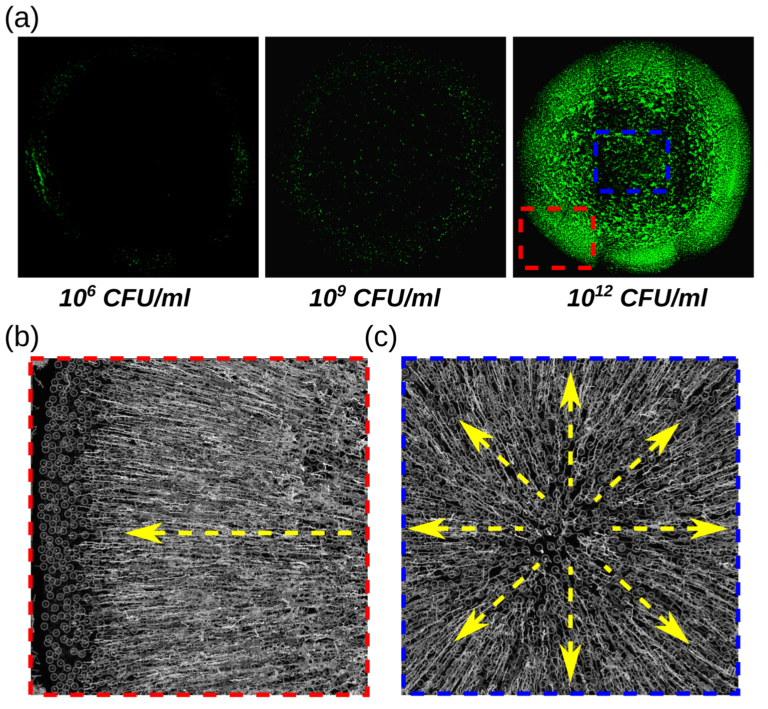}
\centering
\caption{
(a) Confocal microscopy images depicting the final bacterial deposit pattern for $10^6$ CFU/ml, $10^9$ CFU/ml, and $10^{12}$ CFU/ml. (b) Radially outward pathlines of bacteria at the outer edge of the evaporating droplet (refer to red dotted rectangle in (a)) for $10^{12}$ CFU/ml. (c) Radially outward pathlines of bacteria at the center of the evaporating droplet (refer to blue dotted rectangle in (a)) for $10^{12}$ CFU/ml.
}
\label{Figure16}
\end{figure*}

\subsection{Characterization of dried blood residues and bacterial distribution}
Figure 12 (a) shows the SEM (scanning electron microscope) image of the dried residue of whole blood drop. The scale bar in white represents $1.3$ mm. Various kinds of cracks are generally observed, ranging from radial cracks at the thickest portion to mudflat-type cracks at the drop center and close to the droplet's outer edge. The cracks result from desiccation stress during the phase transition from wet to dry gel. Cracks, in general, intersect each other either at right angles or $120^{\circ}$. The resulting cracks fragment the dried residue into several flakes. Figure 12 (b) shows the radial size distribution of the flakes for a typical dried residue of contact radius $R=2.7$mm. Here, $A_f$ denotes the area of a particular flake with a radial centroid coordinate $r_f$, and $R$ denotes the drop contact radius. The flake size is highest between $r_f=1600-2300{\:}{\mu}$m, i.e., in the peripheral region ($r_i=1600{\:}{\mu}$m, $r_0=2300{\:}{\mu}$m). The flake size is smaller in the drop center and the outer edge of the droplet.
Various regions of the dried residue can be characterized following Brutin et al. \citep{brutin2011pattern}.
The region $0<r<r_i$ refers to the central area of the drop, and the residue pattern is formed due to wetting deposits in general. The region $r_i<r<r_0$ forms the outer rim region called the corona. The deposition in the corona region is majorly due to the radially outward transport of RBCs due to the capillary flow inside the evaporating droplet. The outermost region $r_0<r<R$ near the contact line of the drop has a deposit structure and crack morphology similar to the drop center formed due to wetting deposit. Figure 12(c) depicts the various cracks formed due to desiccation stress while transforming from wet gel to dry gel in stage C of blood drop evaporation. Different regions of interest in the dried residue are shown in red rectangles and labeled with 1-4 numerals. The central region labeled as 1 generally has mud-flat-type cracks. The cracks in the central region are typically shaped like ``Y" at a vertex (intersection of cracks). Further, the cracks at a ``Y" vertex typically subtend approximately equal angles close to $120^{\circ}$ 
(figure 12 (c)-1).
The cracks in the corona region labeled as 2 and 3 are typically ``T" shaped and rectilinear. The cracks generally are radial and orthoradial (figure 12 (c)-2), 
intersecting at a right angle at the particular vertex 
(figure 12 (c)-3).
The crack pattern of whole blood is drastically different from the crack pattern formed in pure blood plasma. Refer to supplementary figure S8 for the various types of crack formed in evaporating pure plasma drop.
Figure 13(a) and 13(b) depicts the final precipitate for $10^{9}$ CFU/ml and $10^{12}$ CFU/ml using scanning electron microscope (SEM) respectively. Figure 13(c) shows the probability density function (PDF) $f({\theta})$ for the angle of deviation ${\theta}$ (refer figure 13(b)) from the radial direction. 
The angle ${\theta}$ defined in figure 13(b) measures the degree a crack deviates from the radial direction.
From figure 13(c) it is clearly evident that the crack patterns in the peripheral corona region of the droplet 
deviates from the radial direction significantly for relatively higher bacterial concentration in comparison to lower bacterial concentration. Figure 13(d) shows the radial distribution of flake size represented as an area fraction/percentage for the whole droplet precipitate for $10^{9}$ CFU/ml and $10^{12}$ CFU/ml respectively. Further, it is also evident from figure 13(d) that the coronal flake size are relatively larger for $10^{12}$ CFU/ml bacterial concentration in comparison to $10^{9}$ CFU/ml.
Figure 14 depicts a typical microcharacterization of the dried residue of whole blood using optical profilometry. 
Figure 14(a) shows the 1D thickness profile along the section shown by a black horizontal line (figure 14(b)) passing through the drop diameter. The black dots represents the raw profilometry normalized height data and the red curve is the normalized mean thickness profile of the droplet. Figure 14 (b) depicts the 2D thickness profile filled-contour. Figure 14 (c) depicts a 3D visualization of the surface thickness profile for the dried blood precipitate. Note that the thickness of the dried precipitate is the smallest at the drop center and the outer contact line of the droplet. We observe that the crack thickness, length, and flake size at any given position are proportional to the thickness of the dried residue. Therefore, regions with higher thickness will have larger crack lengths and flake sizes. Physically, the crack length represents a scale over which the cracks form and stresses relax \citep{goehring2013evolving}. 
We also observe that bacteria-laden blood droplets do not show a significant difference in terms of evaporation and dried residue characteristics from pure whole blood droplets within the concentration range typically found in living organisms ${\sim}<(10^9)$ CFU/ml (refer to figure 2(b)). The negligible difference in drop evaporation and dried residue characteristics is probably due to RBCs being one order of magnitude (${\sim}\mathcal{O}(10)$) bigger in length scale and hence approximately two order (${\sim}\mathcal{O}(10^2)$) larger in terms of area ratio.
The maximum bacterial concentration of $10^9$ CFU/ml will give approximately similar number density as RBCs. However, owing to the area ratio, the bacteria will be uniformly distributed throughout the plasma protein matrix with the highest fraction embedded and packed between the biconcave curvatures of the RBCs. For bacteria to have appreciable effects in the corresponding evaporation physics, the 
bacterial number density
has to be significantly altered by the presence of bacteria. 
Alteration of 
bacterial number density
is only possible at very high bacterial concentrations 
($c{\sim}10^{12}$ CFU/ml).
From figure 2(b), we observe that $10^{12}$ CFU/ml reduces the evaporation rate in comparsion to lower bacteria concentration. This reduction in
evaporation rate causes flake size size to be larger compared to high evaporation rate \citep{zeid2013influence}.
At these extreme concentrations of $10^{12}$ CFU/ml the crack pattern in the corona region deviates significantly from that found in lower concentrations (refer to figure 2(d), 13(a), and 13(b) for the difference in crack patterns). Figures 15 (a) and 15 (b)
depict the bacterial deposition in the dried residue of blood plasma and whole blood droplets using confocal fluorescence microscopy. The scale bar in white denotes $500{\:}{\mu}$m. Figure 15
(c-e) shows the RBCs' shape, deposition, and stacking in the dried blood residue using SEM. Generally, the RBCs stack in columnar structures known as rouleaux due to Smoluchowski aggregation/coagulation kinetics \citep{samsel1982kinetics,samsel1984kinetics,barshtein2000kinetics}. Due to the evaporation of blood droplets, the solute concentration increases. As a result, blood tonicity changes from isotonic to hypertonic, causing the RBCs to shrink in size due to water transport out of the red cells, as shown schematically in figure 15(f). The water transport outside the RBCs is caused by the osmotic pressure gradient, resulting in shrunken and wrinkled RBCs inside the precipitate of the evaporating droplet. 
Pal et al. \citep{pal2020concentration} speculates the wrinkled structures to be deformed WBCs. However, owing to the very high number concentration in comparison to WBCs and platelets the authors show that the dominant processes and structure formation that occur during blood droplet evaporation is due to the transport and deposition of RBCs. The ridge-like structures are deformed RBCs that forms during osmosis as explained above. This could also be understood by looking at the relevant length scales involved. It is evident from figure 15(c)-(f) that the ridge like structures forms on top of surfaces which are well structured and have dimensions/length scales similar to RBCs. Also such structures are found in Rouleax stacks which are a well known signature of RBCs stacking. WBCs in general are bigger than RBCs and also does not have a smooth spherical shape. Further WBCs do not show Rouleax kind of stack formation and hence eliminates the possibility of ridge like structures to be deformed WBCs. Platelets are one order smaller than RBCs and hence cannot be related due to ridge like structures as the structures are seen on objects having length scales similar to RBCs.
The distribution of bacteria in the dried blood residue for 
$10^6$ CFU/ml, $10^9$ CFU/ml and $10^{12}$ CFU/ml
can be seen from confocal microscopy images shown in figure 16(a). Figure 16(b) and 16(c) shows fluorescence tagged bacterial path lines from live confocal microscopy at the outer edge and the central region of the evaporating droplet for $10^{12}$ CFU/ml bacterial concentration. Refer to supplementary movie4 - movie9 to visualize the radially outward bacterial motion for various different bacterial concentration and region of interest (drop edge and the center). 

It is clear from figure 16 that the bacteria moves approximately radially outwards on average towards the pinned contact line due to the internal capillary flow generated inside the evaporating droplet. This is also evident from the radially outward bacterial path lines shown in figure 16(b) and 16(c). The actual local trajectory of the individual bacteria are curvilinear and very complex in general as they travel through a multicomponent phase consisting of various cellular components and plasma (refer to supplementary figure S9). The individual trajectories of the bacteria are clearly visible from the live confocal imaging (refer to supplementary movie6 and movie7). The actual local deposition of the bacteria in the final blood precipitate depends on the superposition of the outward radial capillary flow in conjunction to the local perturbation of the bacterial trajectories caused due to the presence of cellular and protein components. The highly irregular cracks in the coronal region for very high bacterial concentration is probably related to the thread like bacterial trajectories that gets packed in the final dry gel precipitate. A thorough understanding of the microphysics and the corresponding fluid dynamics is beyond the scope of the current study and shall be taken as a future endeavour.

\section{Conclusion}
In conclusion, we study the mechanics of sessile whole blood drop evaporation using direct experimental visualization and theoretical methods like time sequence analysis, lubrication analysis, and micro/nano-characterization. We identified that blood drop evaporation, in general, can be subdivided into three stages (A, B, C) based on the evaporation rate. Stage A is the fastest and consists of the gelation of the contact line due to the radial transport of RBCs caused by the outward capillary flow. The gelation occurs due to sol-gel phase transition. Stage B consists of an intermediate evaporation rate in which the gelation front moves radially inward. The radially inward gelation front propagation and droplet height reduction cause the phase transition of the entire droplet into a wet-gel state. The wet-gel phase changes to a dry-gel state in stage C of droplet evaporation. 
Further, we show that the precipitate thickness profile computed from the theoretical analysis conforms to the optical profilometry measurements.
We also observe lamination to delamination transition in stage C which provides quantitative data on the slowest stage C of evaporation. Further, as the wet gel transitions to dry gel, the final dried residue precipitate undergoes desiccation-induced stresses that lead to different kinds of cracks forming on the precipitate. The crack characteristics depend highly on the precipitate's local thickness; hence, a one-to-one map exists between precipitate thickness and the corresponding cracks that are formed. We further show that the drop evaporation rate and final dried precipitate pattern do not change appreciably within the parameter variation of the bacterial concentration
($c{\leq}10^9$ CFU/ml)
typically found in bacterial infection of living organisms. 
However, at very high bacterial concentration ($c{\sim}10^{12}$ CFU/ml) which are biologically not feasible for a living organism we observe the cracks formed in the coronal region of the precipitate deviates from the typical radial cracks found in lower concentrations.



\section*{Supplementary Movie Captions}
Movie 1: Supplementary movie depicting a sample top view. Recorded at 1 frames per 10 seconds, playback speed 10 frames per second.

Movie 2: Supplementary movie depicting a sample side view. Recorded at 1 frames per 10 seconds, playback speed 10 frames per second.

Movie 3: Supplementary movie depicting a sample bottom view. Recorded at 1 frames per 10 seconds, playback speed 10 frames per second.

Movie 4: Supplementary movie depicting radially outward bacterial motion at the edge of the evaporating droplet using live confocal fluorescence microscopy for $10^{6}$ CFU/ml bacterial concentration. Recorded at 1 frame per second, playback speed 10 frames per second.

Movie 5: Supplementary movie depicting radially outward bacterial motion at the edge of the evaporating droplet using live confocal fluorescence microscopy for $10^{9}$ CFU/ml bacterial concentration. Recorded at 1 frame per second, playback speed 30 frames per second.

Movie 6: Supplementary movie depicting radially outward bacterial motion at the edge of the evaporating droplet using live confocal fluorescence microscopy for $10^{12}$ CFU/ml bacterial concentration. Recorded at 1 frame per second, playback speed 30 frames per second.

Movie 7: Supplementary movie depicting radially outward bacterial motion at the center of the evaporating droplet using live confocal fluorescence microscopy for $10^{12}$ CFU/ml bacterial concentration. Recorded at 1 frame per second, playback speed 30 frames per second.

Movie 8: Supplementary movie depicting radially outward bacteria pathlines at the edge of the evaporating droplet using live confocal fluorescence microscopy for $10^{12}$ CFU/ml bacterial concentration. Recorded at 1 frame per second, playback speed 30 frames per second.

Movie9: Supplementary movie depicting radially outward bacteria pathlines at the center of the evaporating droplet using live confocal fluorescence microscopy for $10^{12}$ CFU/ml bacterial concentration. Recorded at 1 frame per second, playback speed 30 frames per second.

\section*{Declaration of Interests}
The authors declare no conflict of interest.
\section*{Acknowledgement}
The authors acknowledge and thank Navin Kumar Chandra for helping with the viscosity measurements of the blood samples.
\clearpage
\bibliographystyle{jfm}
\bibliography{jfm-instructions}

\end{document}